\pdfoutput=1
\documentclass{emulateapj}  
\usepackage{courier,graphicx,epsfig,color,colordvi,natbib,url,twoopt}
\usepackage[breaklinks=true]{hyperref} 

\bibpunct{(}{)}{;}{a}{}{,}

\def\deg{\hbox{$^\circ$}}      
\def\mAA{m\AA}
\def\Halpha{\mbox{H\hspace{0.1ex}$\alpha$}} 
\def\FeI{\ion{Fe}{1}\,6301\,\AA}
\def\CIV{\ion{C}{4}}
\def\CaIIH{\ion{Ca}{2}\,H}
\def\CaII{\ion{Ca}{2}\,8542\,\AA}
\def\CaIR{\ion{Ca}{2}}
\def\HeII{\ion{He}{2}\,304\,\AA}
\def\FeIX{\ion{Fe}{9}\,171\,\AA}
\def\FeXIV{\ion{Fe}{14}\,211\,\AA}
\def\etal{et al.}          
\def\EB{Ellerman bomb}
\def\EBs{Ellerman bombs}
\def\is{\!=\!}
\def\eg{e.g.,}
\def\ie{i.e.,}

\long\def\rev#1{#1}           
\long\def\moved#1{#1}           
\long\def\revv#1{\textbf{#1}}   
\long\def\revv#1{#1}           

\makeatletter
 \newcommandtwoopt{\citeads}[3][][]{\href{http://adsabs.harvard.edu/abs/#3}%
   {\def\hyper@linkstart##1##2{}%
    \let\hyper@linkend\@empty\citealp[#1][#2]{#3}}}
 \newcommandtwoopt{\citepads}[3][][]{\href{http://adsabs.harvard.edu/abs/#3}%
   {\def\hyper@linkstart##1##2{}%
    \let\hyper@linkend\@empty\citep[#1][#2]{#3}}}
 \newcommandtwoopt{\citetads}[3][][]{\href{http://adsabs.harvard.edu/abs/#3}%
   {\def\hyper@linkstart##1##2{}%
    \let\hyper@linkend\@empty\citet[#1][#2]{#3}}}
 \newcommandtwoopt{\citeyearads}[3][][]%
   {\href{http://adsabs.harvard.edu/abs/#3}%
   {\def\hyper@linkstart##1##2{}%
    \let\hyper@linkend\@empty\citeyear[#1][#2]{#3}}}
\makeatother
  
\newcommand{\gregalemail}{g.j.m.vissers@astro.uio.no}

\begin{document}

\title{Ellerman bombs at high resolution: \\  
II. Triggering, visibility, and effect on upper atmosphere}
 
\author{Gregal J. M. Vissers}
\author{Luc H. M. Rouppe van der Voort}
\author{Robert J. Rutten}
\affil{Institute of Theoretical Astrophysics,
  University of Oslo, %
  P.O. Box 1029 Blindern, N-0315 Oslo, Norway; \gregalemail}

\shorttitle{Ellerman bombs at high resolution II}
\shortauthors{Vissers et al.}

\begin{abstract}
  We use high-resolution imaging spectroscopy with the Swedish 1-m
  Solar Telescope (SST) to study the transient brightenings of the
  wings of the Balmer \Halpha\ line in emerging active regions that
  are called Ellerman bombs.
  Simultaneous sampling of \CaII\ with the SST confirms that most
  \EBs\ occur also in the wings of this line, but with markedly
  different morphology.
  Simultaneous images from the Solar Dynamics Observatory (SDO) show
  that \EBs\ are also detectable in the photospheric 1700\,\AA\
  continuum, again with differing morphology.
  They are also observable in 1600\,\AA\ SDO images, but with much
  contamination from \CIV\ emission in transition-region features.
  Simultaneous SST spectropolarimetry in \FeI\ shows that \EBs\ occur
  at sites of strong-field magnetic flux cancelation between small
  bipolar strong-field patches that rapidly move together over the solar
  surface.
  Simultaneous SDO images in \HeII, \FeIX, and \FeXIV\ show no clear
  effect of the \EBs\ on the overlying transition region and corona.
  These results strengthen our earlier suggestion, based on \Halpha\
  morphology alone, that the \EB\ phenomenon is a purely photospheric
  reconnection phenomenon.
 \end{abstract}

\keywords{Sun: activity -- Sun: atmosphere -- Sun: magnetic topology}

\section{Introduction}\label{sec:introduction}
\moved{\EBs\ are transient brightenings of the outer wings of the Balmer
\Halpha\ line at 6563\,\AA\ that occur in solar active regions with
much flux emergence
(\citeads{1917ApJ....46..298E}). 
The fairly extended literature on this topic was summarized by
\citetads{2002ApJ...575..506G} 
and reviewed more recently by
\citetads{2013JPhCS.440a2007R}. 

Ellerman bombs are of particular interest because they seem to
pinpoint emerging magnetic field.
Various topologies have been proposed: reconnection between emerging
flux and existing fields
(\citeads{2008ApJ...684..736W}; 
\citeads{2010PASJ...62..879H}; 
\citeads{2010PASJ...62..901M}), 
reconnection between shearing unipolar fields
(\citeads{2002ApJ...575..506G}; 
\citeads{2008ApJ...684..736W}; 
\citeads{2010PASJ...62..879H}), 
and a much-elaborated scenario of reconnection between opposite walls
of $\cup$-shaped fields in undulatory (``sea serpent'') flux emergence
(\citeads{2004ApJ...614.1099P}; 
\citeads{2006AdSpR..38..902P}; 
\citeads{2008ApJ...684..736W}; 
\citeads{2012ASPC..455..177P}; 
\citeads{2012EAS....55..115P}) 
which started with the Flare Genesis Experiment
(\citeads{2002SoPh..209..119B}; 
\citeads{2002ApJ...575..506G}; 
\citeads{2004ApJ...601..530S}). 
The concept of undulatory field emergence with reconnection in the low
atmosphere has also been studied numerically by
\citetads{1992ApJS...78..267N}, 
\citetads{1995Natur.375...42Y}, 
\citetads{1999ApJ...515..435L}, 
\citetads{2007ApJ...657L..53I}, 
\citetads{2008ApJ...687.1373C}, 
and \citetads{2009A&A...508.1469A}. 

In this paper we study \EBs\ using new imaging spectroscopy with the
the Swedish 1-m Solar Telescope (SST;
\citeads{2003SPIE.4853..341S}). 
Its field of view and the typical sequence duration permitted by
atmospheric seeing in ground-based observing make such data less suited
to study \EB\ occurrence as indicator of large-scale active region
emergence patterning, but the unprecedented spatial, temporal, and
spectral resolution of SST data permits microscopic study of the
structure and dynamics of individual \EBs\ with much higher data
quality than all earlier studies.
This paper is a sequel to Watanabe et al.\
(\citeyearads{2011ApJ...736...71W}, 
henceforth Paper~I) who established from such SST data that Ellerman
bombs appear as small, rapidly varying, upright ``flames'' of bright
emission in the \Halpha\ wings that are rooted in magnetic
concentrations.
\revv{These authors concluded that the \EBs} constitute a purely photospheric phenomenon and are hidden
at \Halpha\ line center by overlying chromospheric fibrils.
This morphology suggested heating from strong-field magnetic
reconnection taking place in the low photosphere, not in the chromosphere
as thought so far.

In this sequel we analyze two new SST \Halpha\ data sets, one with simultaneous
\CaII\ imaging spectroscopy, the other with simultaneous \FeI\ imaging polarimetry.
We also add comparison with ultraviolet imaging in the 1600\,\AA,
1700\,\AA, 304\,\AA, 171\,\AA, and 211\,\AA\ passbands of the Atmospheric Imaging Assembly (AIA;
\citeads{2012SoPh..275...17L}) 
on the Solar Dynamics Observatory (SDO).
We use these data to broaden the evidence that \EBs\ mark strong-field
reconnection, to compare \EB\ morphology at high resolution in
\Halpha\ and \CaII, to discuss how to best detect \Halpha\ \EBs\ in
ultraviolet AIA images so that the huge AIA database may become
available for \EB\ pattern research, and to test claims that \EBs\ are
related to upper-atmosphere phenomena such as surges
and micro-flares.

Our combined \Halpha\ and \FeI\ imaging spectroscopy may be
regarded as higher-resolution analysis of the type as the Hinode
analysis by
\citetads{2008PASJ...60..577M}, 
while our comparison of \Halpha\ and \CaII\ imaging spectroscopy follows on
similar but lower-resolution comparisons by
\citetads{2006SoPh..235...75S}, 
\citetads{2006ApJ...643.1325F}, 
and \citetads{2007A&A...473..279P}. 
There are many reports on \EB\ appearance in 1600~\AA\ TRACE
(\citeads{2000ApJ...544L.157Q}; 
\citeads{2002ApJ...575..506G}; 
\citeads{2006ApJ...643.1325F}; 
\citeads{2006SoPh..235...75S}; 
\citeads{2007A&A...473..279P}; 
\citeads{2007ASPC..368..253P}; 
\citeads{2008PASJ...60..577M}; 
\citeads{2010MmSAI..81..646B}; 
\citeads{2011CEAB...35..181H}), 
but none yet on comparison with AIA's 1700~\AA\ imaging which seems a
better \EB\ diagnostic than its 1600~\AA\ imaging.

Finally, there are reports of upper atmosphere response to underlying
\EBs\ in the form of \Halpha\ surges
(\citeads{1973SoPh...28...95R}; 
\citeads{1973SoPh...30..449R}; 
\citeads{1982SoPh...77..121S}; 
\citeads{2008PASJ...60...95M}; 
\citeads{2010ApJ...724.1083G}; 
Paper~I) 
but the ubiquity of such correspondence seems questionable
(Paper~I). 
The same holds for correspondence between \EBs\ and energetic
upper-atmosphere phenomena 
(\citeads{2002ApJ...574.1074S}; 
\citeads{2009ApJ...701..253M}) 
of which the ubiquity was also questioned by
\citetads{2004ApJ...601..530S} 
who found only one Flare Genesis example amidst hundreds of \EBs.
The availability of better-quality short-wavelength imaging with SDO
warrants and enables further investigation of such correspondences.

The structure of the paper is as follows.
In Sect.~\ref{sec:observations} we describe the observational
procedures and the data.
The analysis methods are explained in Sect.~\ref{sec:analysis}.
The results are presented in Sect.~\ref{sec:results} and discussed in
Sect.~\ref{sec:discussion}.
We end the paper with conclusions in Sect.~\ref{sec:conclusion}.
}

\section{Observations and Data Reduction}\label{sec:observations}
\subsection{SST/CRISP data acquisition and reduction}
\paragraph{Observational setup}
As in Paper~I, we use data obtained with the CRisp Imaging
SpectroPolarimeter (CRISP;
\citeads{2008ApJ...689L..69S}) 
at the SST.
CRISP and the SST together yield imaging spectropolarimetry at
unsurpassed high spatial, spectral and temporal resolution. 
The telescope is equipped with a real-time tip-tilt and
adaptive-optics wave-front correction system
(\citeads{2003SPIE.4853..370S}). 
CRISP is a dual Fabry-P\'erot interferometer (FPI) operating in the
red part of the spectrum that allows wavelength tuning within 50\,ms.
The light from the telescope is first guided through an optical
chopper which ensures strict synchronization of exposures by three
cameras.
The wavelength band is selected with a prefilter mounted on a
filterwheel that allows a spectral band change within 250--600\,ms. 
CRISP contains two liquid crystals for polarimetry and high-resolution
and low-resolution etalons for wavelength selection and tuning.
The polarimetric modulation is accomplished by cycling the liquid
crystal variable retarders through four different states. 
An orthogonally polarizing beam splitter behind the FPI divides the
light onto two cameras in order to reduce seeing-induced cross-talk
(cf.~\citeads{1987ApOpt..26.3838L}). 
Between the prefilter and CRISP a few percent of the light is branched
off to a camera imaging this wide band to serve as ``multi-object''
anchor in the post-processing.
The three CCD cameras are identical high-speed low-noise Sarnov
CAM1M100 cameras with 1K$\times$1K chips. 
They run at 35\,fps frame rate with an exposure time of 17\,ms. 

\paragraph{Data acquisition and reduction}
Two SST/CRISP data sets are used in this study.
The first was acquired on 2010 June 28 during 08:16--09:06\,UT
covering 54\arcsec$\times$53\arcsec\ of active region NOAA\,11084 containing a sunspot located at
$(X,Y)=(-720,-343)$ in standard heliocentric solar coordinates (in
arcsec, with the $Y$-direction positive northward and the
$X$-direction positive westward from the center of the apparent solar
disk).
Both the \Halpha\ and \CaII\ profiles were finely sampled in this
observation.
Full Stokes data were intended to be taken in the \FeI\ line at
$-$48\,m\AA\ but unfortunately, the wavelength setting was incorrect;
these data are not used in this study.

The second data set was obtained on 2011 May 7 during
08:56--09:52\,UT, with the field of view centered on
$(X,Y)=(317,306)$, 
covering 55\arcsec$\times$55\arcsec\ and containing part of a sunspot
and some pores in active region NOAA\,11204.
In this observation, \Halpha\ was sampled only at $\Delta \lambda \is
\pm 1$\,\AA\ and at line center.
The profiles of \CaII\ and \FeI\ were finely sampled including full
Stokes polarimetry but we do not use the \CaIR\ data in this study.
Further detail including viewing angles, spectral passbands,
wavelength samplings, cadences, and durations for both data sets are
given in Table~\ref{tab:datasets}.

\begin{table*}[bth]
\caption{Overview of the CRISP data sets analyzed in this study.}
\begin{center}
\begin{tabular}{ccrclcccccc}%
	\hline \hline
	{}    & {}      & {}	  & {}	      & {}          & Prefilter & CRISP & {}    & {}        & {} &
  {} \\
	{}	& {}		  & {}    & Location  & {}          & passband  & FWHM  & Range	& Sampling  &
  Cadence		&	Duration \\
	Data set		& Target  & Date  & ($\mu$)   & Diagnostic  & [\AA]     & [\mAA]& [\AA]	& [\mAA]	  & [s]		&	[min] \\
	\hline
	1	  & AR~11084	&	2010 Jun 28	& \rev{0.53}  & \Halpha~6563\,\AA & 4.9 & 66 & $\pm$1.9 	& 85	&  22.4	& 51 \\
	{}  & {}      	&	{}          & {}    & \CaII\       & 9.3 & 111 & $\pm$1 	& 55	&  {}	& {} \\
	\hline
	2	  & AR~11204	&	2011 May 7	& \rev{0.89} & \Halpha~6563\,\AA & 4.9 & 66 & $\pm$1 	& 1000	&  27.4	& 56 \\
	{}  & {}      	&	{}          & {}    & \FeI\         & 4.6 & 64 & $-$0.6--1.7 	& 48$^{\rm{a}}$ & {} \\
	\hline
\end{tabular}
\begin{minipage}{.9\hsize}
  $^{\rm{a}}$ The indicated spacing holds between $-$480 and +576\,m\AA\ but two extra continuum 
  samplings were added at $-$610\,m\AA\ and +1734\,m\AA.
\end{minipage}
\end{center}
\label{tab:datasets}
\end{table*}%

At each \Halpha\ and \CaII\ wavelength sampling a ``multi-frame''
burst of eight exposures was taken.
For \FeI\ in the second data set four exposures were recorded for each
liquid-crystal state at each wavelength position.
The image scale is 0\farcs0592\,px$^{-1}$, well below the SST's
Rayleigh diffraction limit for \Halpha\ (0\farcs17), \CaII\
(0\farcs21), and \FeI\ (0\farcs16).
  
Post-processing with Multi-Object Multi-Frame Blind Deconvolution
(MOMFBD, \citeads{2005SoPh..228..191V}) 
reduced the remaining high-order image deterioration from atmospheric
seeing considerably.
In this technique, all images at each tuning position within a line
profile scan are tessellated into $64\times64$\,px$^2$ overlapping
subfields that are each MOMFBD-restored independently and then
re-assembled. 
The wide-band images act both as multi-object channel for numerical
wavefront sensing and as alignment anchor for the narrow-band CRISP
images.

Remaining small-scale seeing deformations due to the non-simultaneity
of the sequentially tuned narrowband CRISP images are minimized by
application of the cross-correlation method of
\citetads{2012A&A...548A.114H}. 
The data are subsequently also corrected for the transmission profile
of the prefilter following \citetads{2012PhDT.........8D}. 

The final post-processing of the image sequences includes correction
for the time-dependent image rotation that results from the
alt-azimuth configuration of the SST, and destretching following
\citetads{1994ApJ...430..413S} 
which removes remaining rubber-sheet distortions.
The latter are determined from the wide-band images and then applied
to the co-aligned narrow-band ones.

The polarimetric \FeI\ data were processed following
\citetads{2012ApJ...757...49W}, 
which is a modification of the method developed by
\citetads{2011A&A...534A..45S}. 

\subsection{SDO/AIA data reduction and co-alignment}
\begin{figure*}[bht]
  \centerline{\includegraphics[width=\textwidth]{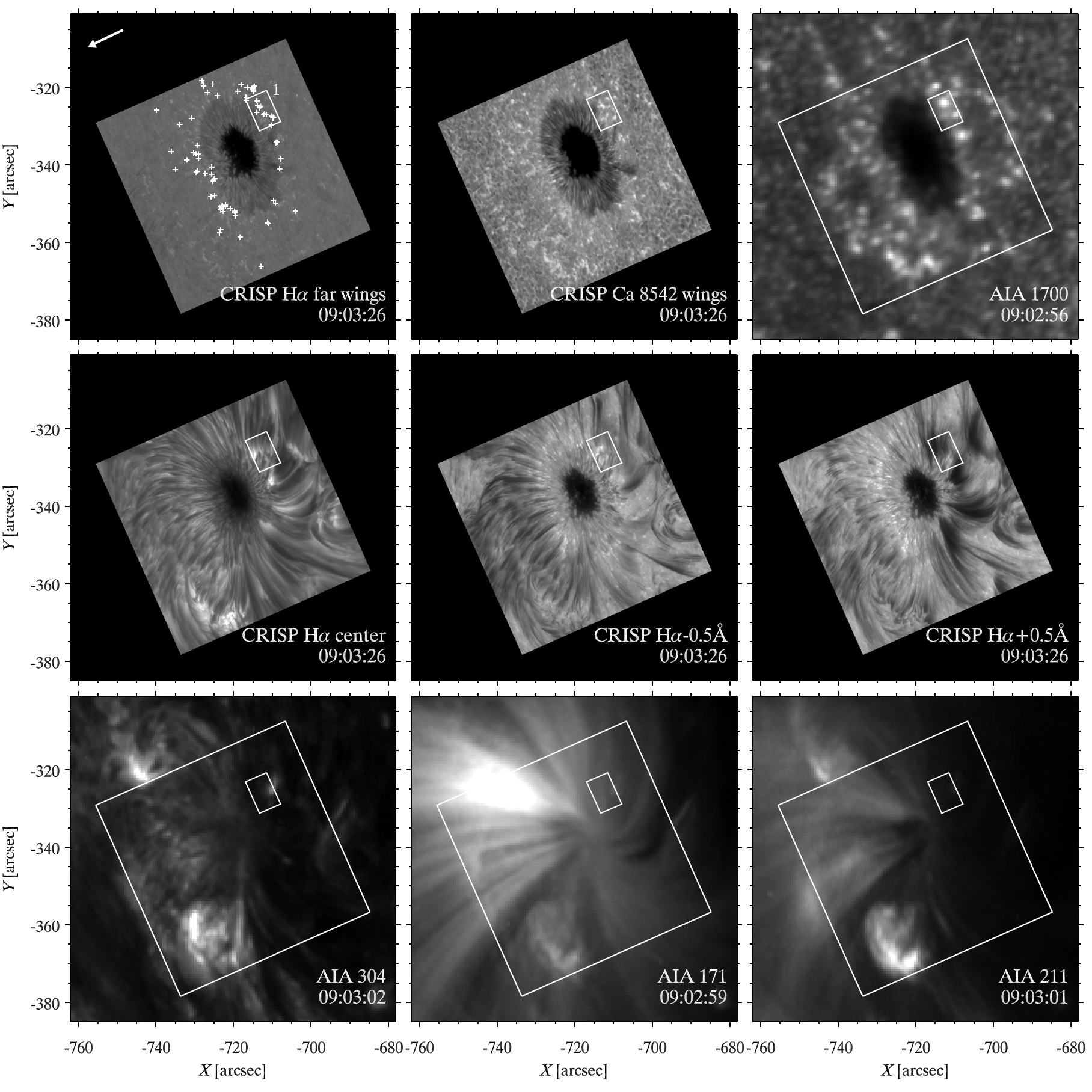}}
  \caption[]{\label{fig:fov1} %
    Near-simultaneous co-aligned CRISP and AIA image samples from
    data set 1.
    The SST field of view is rotated to heliocentric $(X,Y)$
    coordinates and is specified by a large white square in the AIA
    panels.   
    In the first panel the arrow specifies the direction towards the
    limb and the plus signs specify the locations of all \EBs\
    detected in data set 1.
    Each image is scaled independently.
    The small rectangle marks cutout region-of-interest 1, containing
    a bright \EB\ at this time which is also seen in the second and
    third panels.
    {\it First row\/}: photospheric diagnostics \Halpha\ summed wing ($\pm$(0.9--1.1)\,\AA)
    intensity, \CaII\ summed wing intensity ($\pm$(0.6--0.7\,\AA)), 1700\,\AA\ intensity.
    {\it Second row\/}: chromospheric diagnostics \Halpha\ line center
    intensity, \Halpha\ blue and red wing intensities at $\Delta
    \lambda \is \pm 0.5$~\AA.
    {\it Third row\/}: transition region diagnostics 304\,\AA,
    171\,\AA\ and 211\,\AA\ intensities.
  }
\end{figure*}

\begin{figure*}[bhtp]
 \centerline{\includegraphics[width=\textwidth]{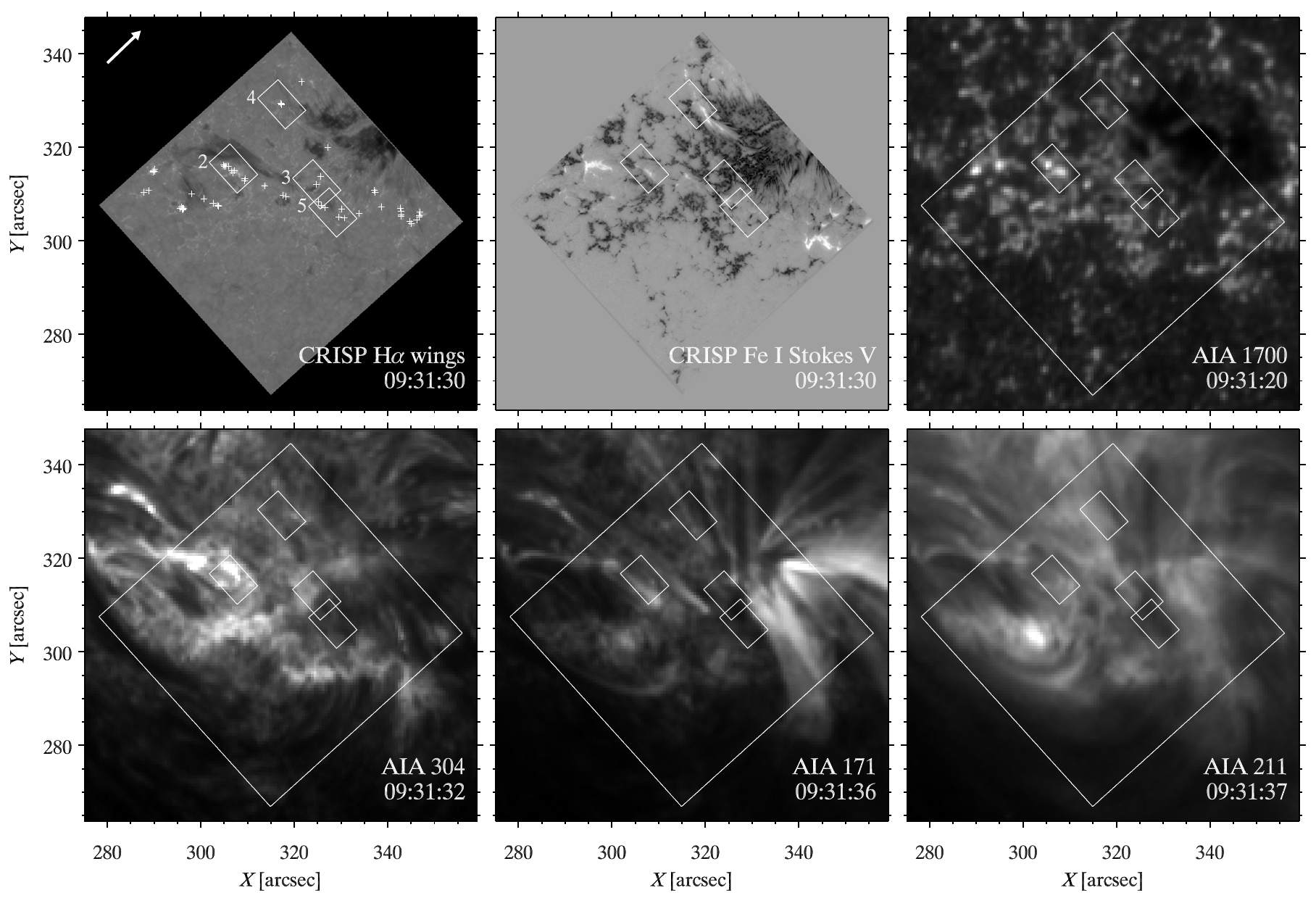}}
 \caption[]{\label{fig:fov2} %
   Near-simultaneous co-aligned CRISP and AIA image samples from
   data set 2 in the format of Fig.~\ref{fig:fov1}.
   The small rectangles mark the cutouts of regions of interest 2--5.
   {\it Upper row\/}: photospheric diagnostics \Halpha\ summed wing
   intensity ($\pm$1\,\AA), \FeI\ Stokes-$V/I$, 1700\,\AA\ intensity.
   {\it Lower row\/}: transition region diagnostics 304\,\AA,
   171\,\AA\ and 211\,\AA\ intensities.
 }
\end{figure*}

For both SST data sets we selected overlapping SDO/AIA images in the
1600\,\AA, 1700\,\AA, 304\,\AA, 171\,\AA, and 211\,\AA\ wavelength
channels, covering 84\arcsec$\times$84\arcsec\ and centered on the
field of view of the SST.
The level-1 AIA data were improved to level-1.5 with the SolarSoft
\texttt{aia\_prep.pro} procedure, yielding data with spatial sampling
of 0\farcs6\,px$^{-1}$, a temporal cadence of 24\,s for the 1600\,\AA\
and 1700\,\AA\ data, and 12\,s for the other channels.
The SST images were co-aligned to the AIA images using bright points
in the blue wing of \Halpha\ and in the 1700\,\AA\ images as reference
for cross-correlation, taking the SDO image closest in time per SST
image.
Figure~\ref{fig:fov1} shows image samples from data set 1;
Figure~\ref{fig:fov2} shows image samples from data set 2.

\section{Analysis methods} \label{sec:analysis}
\subsection{\EB\ detection and selection}\label{sec:detection}
In Paper~I \EBs\ were identified manually on the basis of their
flame-like morphology.
In this paper we developed an automated detection algorithm through
extensive trials in which visual inspection showed which constraints
work best to properly recover the \EB\ flames seen in our data.
The algorithm consists of four constraints:

\begin{enumerate}

\item {\it Brightness.}
  A double intensity threshold is applied to the SST \Halpha\ and
  \CaII\ data.
  First, only pixels exceeding a threshold of \rev{155\% of} the
  average intensity over the whole field of view are passed.
  Second, a lower threshold of \rev{140\% of} the average then
  passes only those pixels that are adjacent to already selected ones.
  
\item {\it Size.}
  To emulate the visually observed flame morphology, we require that
  at least five of the selected pixels are spatially connected,
  corresponding to 0\farcs2--0\farcs3 extent.

\item {\it Continuity.}
  Detections meeting the above constraints in subsequent images often
  overlap spatially and are then considered to represent the same
  event.
  However, sometimes temporal gaps occur due to variable seeing.
  We therefore allow that detections may skip up to two frames ($\sim$50\,s) before
  overlapping again to still be counted as a single event.
  Also, splitting or merging events are resolved at this stage by
  propagating the detection with the largest spatial overlap between
  frames, while the one with the smallest overlap is considered to
  originate, respectively disappear, at that particular time step.
  
\item {\it Lifetime.} 
  Finally, all detections that are visible for less than two
  consecutive images (corresponding to $\sim$45\,s and $\sim$55\,s for
  data set 1 and 2, respectively) are removed from the sample. 
  Note that this duration threshold differs considerably from the
  value of 240\,s used in Paper~I.
  It results from trade-off between maximizing the number of detected
  \EBs\ and reducing the number of false identifications.
  
\end{enumerate}

Summed \Halpha\ wing data (obtained by taking the average of
\Halpha$\pm$(0.9--1.1)\,\AA\ and \Halpha$\pm$1.1\,\AA\ for data set 1
and 2, respectively), as well as \CaII\ summed blue and red wing images
(obtained by taking the average of three wing positions covering $\pm$(0.6--0.7)\,\AA\ in either
wing separately) were run through this detection algorithm.
The selected \CaIR\ wing positions were chosen as such to minimize the effects of overlying fibril
obscuration.
For comparison purposes, the algorithm was also tested with a single 5-$\sigma$ above average brightness
threshold on the 1700\,\AA\ data.
Table~\ref{tab:detstats} gives an overview of the detection results
after steps 2, 3, and 4 (with the corresponding number of remaining
detections in the third through fifth columns, respectively).

\begin{table}[h]
\caption{\revv{Number statistics from automated detection.}}
\begin{center}
\begin{tabular}{llrrr}%
	\hline \hline
  Set  & Diagnostic  &  \multicolumn{3}{c}{\revv{Number of d}etections after threshold} \\
  {}        & {}          & Int.~\& size & Continuity & Lifetime \\
	\hline
  1         & \Halpha~6563\,\AA       & \rev{783}         & \rev{106}         & \rev{78} \\
  {}        & \CaII\ total$^{a}$      & ---         & ---         & \rev{13} \\
  {}        & {\it -- \CaII\ blue\/}  & {\it \rev{174}\/} & {\it \rev{18}\/}  & {\it 9\/} \\
  {}        & {\it -- \CaII\ red\/}   & {\it \rev{137}\/} & {\it \rev{14}\/}  & {\it \rev{13}\/} \\
  {}        & Cont.~1700\,\AA         & \rev{294}         & \rev{29}          & \rev{25} \\
  \hline
  2         & \Halpha~6563\,\AA       & \rev{436}         & \rev{81}         & \rev{61} \\
  {}        & Cont.~1700\,\AA         & \rev{420}         & 37          & 32 \\
	\hline
\end{tabular}
\begin{minipage}{.9\hsize}
  $^{a}$ Result of combining the detections in both wings of the line
  and considering spatially overlapping detections at a particular
  time step to represent the same event.
\end{minipage} 
\end{center}
\label{tab:detstats}
\end{table}

It should be noted that the number of actual \EBs\ is probably higher
than suggested by the last column in this table, as a number of
detections displays substructure, also sequentially in time, that is
not differentiated into separate detections by the algorithm.

Our comparisons between the results of our algorithm tests and the
visual appearance of the \EB\ flames in our data was largely done by
extensive use of the CRisp SPectral EXplorer of
(CRISPEX; 
\citeads{2012ApJ...750...22V}) 
of which the browsing and analysis functionality allows fast
confirmation of algorithmic \EB\ detections as well as simultaneous
multi-diagnostic comparisons of multiple \EB\ signatures.

\subsection{Spectropolarimetric analysis}
We investigated the \EB\ behavior in our data with respect to the
magnetic field distribution over the surface both qualitatively and
quantitatively by comparing summed \Halpha\ wing and \FeI\
Stokes-$V/I$ intensity images with the goal to establish a connection
between the \EB\ phenomenon and the underlying magnetic and flow
fields.
Firstly, we determined the distance to the nearest opposite polarity
in the Stokes-$V/I$ image at $-$48\,\mAA\ from line center for every
pixel at every time step.
Secondly, we derived the photospheric surface flow field from the
\FeI\ continuum images at +1734\,\mAA\ using the local correlation
tracking technique of \citetads{1995A&A...295..199Y}, 
applying a temporal window of 4 minutes and Gaussian spatial smoothing
with a halfwidth of 0\farcs7.

\section{Results} \label{sec:results}

\paragraph{Detection statistics}
\rev{When applied to the \Halpha\ data, our algorithm detects 78 and 61 \EBs\ in the first and
second data set, respectively.
Although the longest detections in data set 1 and 2 last about 35\,min and 20\,min, 
the detection lifetimes are on average 3.5--4\,min and at least 75\% of the detections has a
lifetime shorter than 5\,min.
For the 1700\,\AA\ continuum the lifetimes are longer on average, but the lifetime of the longest
living detections is similar to those observed in \Halpha, with at least 70\% of the detections
having a lifetime of 5\,min or shorter.}

\rev{
The average area covered by single detections in the \Halpha\ images is 0.2--0.3\,arcsec$^{2}$
and more than 90\% has an area smaller than 0.6\,arcsec$^{2}$.
The detection sizes in the AIA data are larger, with an average of approximately
1.1--1.3\,arcsec$^{2}$.}

\rev{
For \CaII\ the number of detections is too small to give meaningful statistics, but the results
would suggest they have a similar tendency as the 1700\,\AA\ detections, \ie\ longer lifetimes and
larger area than in \Halpha\ (although the values are much closer to those of \Halpha\ than to those
of 1700\,\AA).
}

\paragraph{Signature in \Halpha\ and \CaII}

\begin{figure*}[h]
 \centerline{\includegraphics[width=\textwidth]{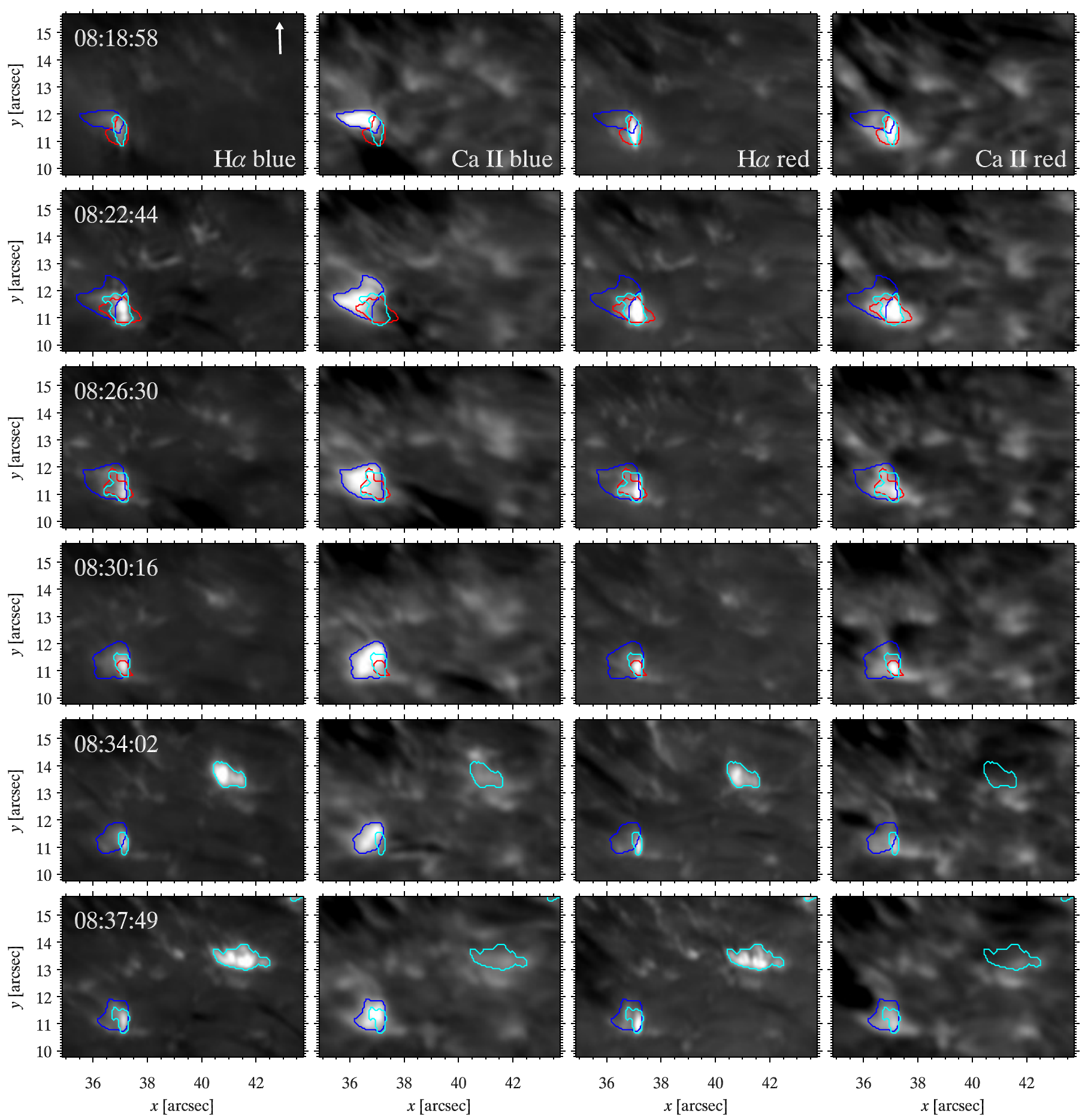}}
 \caption[]{\label{fig:ha-ca-timeseq}%
   \EB\ evolution in a sequence of cutouts in \Halpha\ blue wing
   (\emph{first column\/}), \CaII\ blue wing (\emph{second column\/}), \Halpha\ red wing
   (\emph{third column\/}), and \CaII\ red wing (\emph{fourth column\/})
   for region-of-interest 1 on the center-side of the sunspot in data
   set 1 (cf.~Fig.~\ref{fig:fov1}).
   The cutouts are rotated clockwise by 114\deg\ from their
   orientation in Fig.~\ref{fig:fov2} in order to obtain a close to
   vertical limbward direction which is indicated by the white arrow
   in the top left panel.
   The time in UT is specified in the upper-left corners of the
   panels in the first column.
   In all panels the detection contours based on \Halpha\
   (\emph{azure\/}) and \CaIR\ blue and red wing (\emph{blue\/} and
   \emph{red\/}, respectively) have been overlaid.
 }
\end{figure*}

\begin{figure*}[ht]
 \centerline{\includegraphics[width=\textwidth]{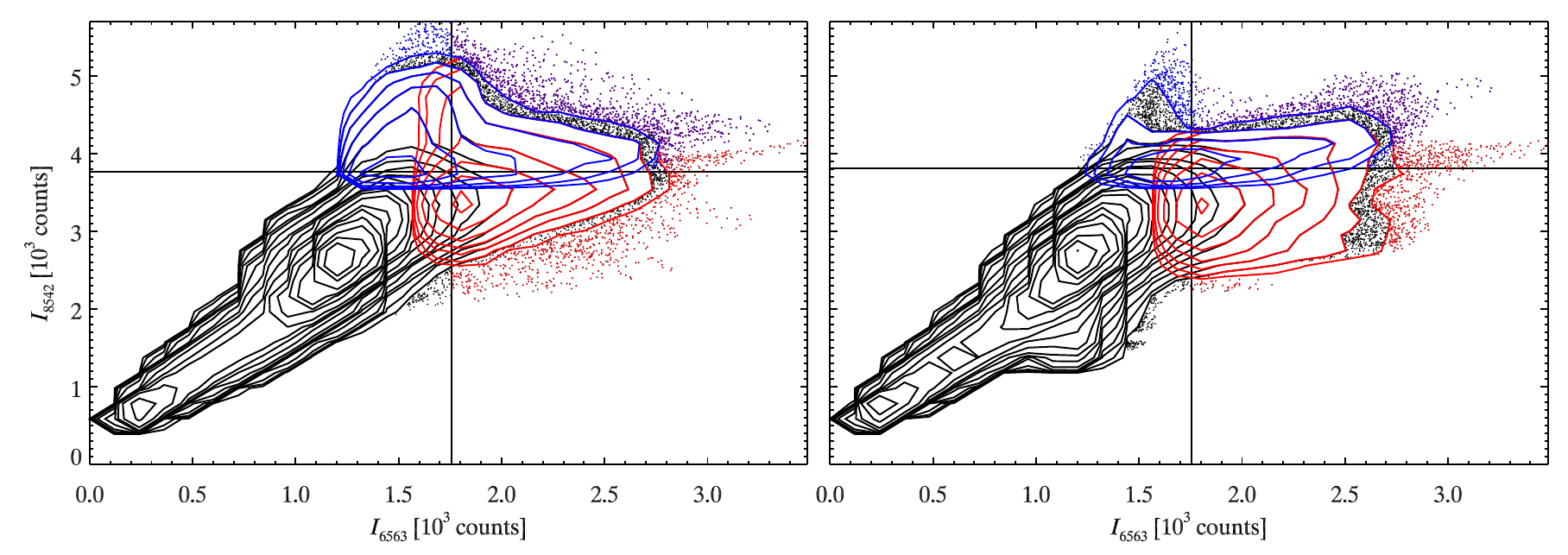}}
 \caption[]{\label{fig:ha-ca-scatter}%
   Scatter diagrams for \Halpha\ versus \CaII\ blue wing (\emph{left
     panel\/}) and red wing (\emph{right panel\/}) for data set 1,
   with contours and points for all pixels (\emph{black\/}), \Halpha\
   detection pixels (\emph{red\/}) and \CaIR\ detection pixels
   (\emph{blue\/}).
   Sample density contours are plotted where high sample numbers occur
   to avoid plot saturation.
   Pixels outside the contours that are common to both \Halpha\ and \CaIR\
   detections are correspondingly purple.
   The vertical and horizontal lines specify the \rev{140\%}
   thresholds of the average intensity over the field of view for
   \Halpha\ and \CaIR, respectively.
 }
\end{figure*}

Figure~\ref{fig:ha-ca-timeseq} shows the time evolution in both
\Halpha\ and \CaII\ of a few selected \EBs\ in data set 1.
Detection contours based on both spectral diagnostics are overlaid on
the images.

The \Halpha\ panels (first and third columns) 
illustrate the basic \EB\ morphology reported in Paper~I, \ie\ they
appear as slender features, upright in the general direction of the
limb.
This is also shown by the azure \Halpha-based detection contours.
However, they are quite variable in both shape and intensity in their
temporal evolution.
Note that not all \Halpha\ detections have a corresponding detection
in \CaIR\ (\eg\ the azure contour in the upper right of the lower eight panels).
Although the shapes of the detection contours based on the \Halpha\
and \CaIR\ images, respectively, often overlap, this overlap is
typically not one-to-one (cf.\ the differences between the detection
contours in the upper four rows).
Also, the detections in the blue and red wings of \CaIR\ are usually
quite dissimilar, as shown by the corresponding blue and red contours
in the upper rows of Fig.~\ref{fig:ha-ca-timeseq}.

Figure~\ref{fig:ha-ca-scatter} quantifies these observations in the
form of scatter diagrams.
For the majority of pixels in this data set, and in particular for
those with an intensity below a \rev{140\% of} average cutoff in
either diagnostic (\ie\ the lower left quadrant), there is a tight
correlation between the \Halpha\ and \CaIR\ intensities.
It continues to larger brightness values in both spectral lines
regardless whether the blue or red wing of \CaIR\ is considered,
although clearest in the latter.
Furthermore, above both thresholds (upper right quadrant) most bright
pixels are detected as \EBs\ in both diagnostics, but some of the
brighter \Halpha-detected pixels above the \CaIR\ threshold are not
detected as such in either \CaIR\ wing.
Conversely, a considerable number of the brighter \CaIR\ pixels falls
below the \Halpha\ threshold (\ie\ the high \CaIR-intensity peak near
the \Halpha\ threshold and in the upper left quadrant).
Also, and in contrast to the blue wing of \CaIR, there are relatively
more pixels in the \CaIR\ red wing that are bright in \Halpha\ but no
so much in \CaIR, \eg\ the ``lump'' in the black contours around
$I_{6563}\is\ $1200\,counts and the more extended contours above the
\Halpha\ and below the \CaIR\ thresholds in the right-hand panel.

\begin{figure}[bht]
 \centerline{\includegraphics[width=\columnwidth]{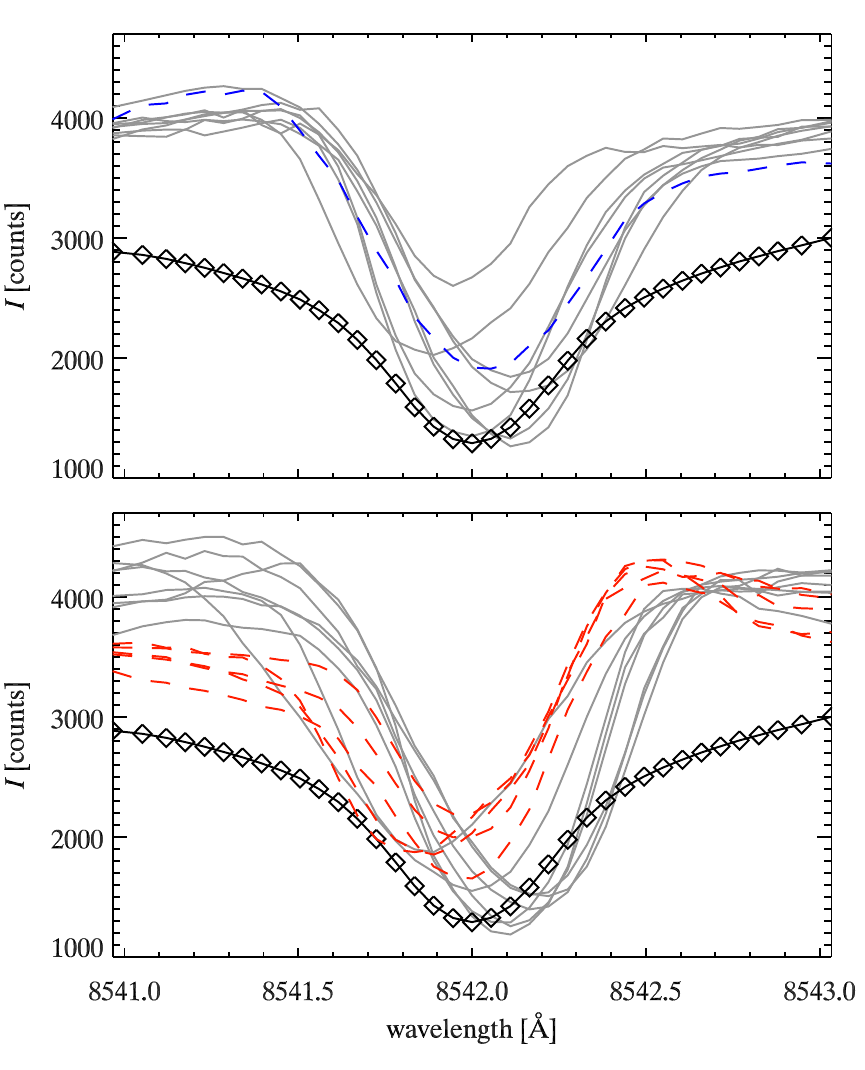}}
 \centerline{\includegraphics[width=\columnwidth]{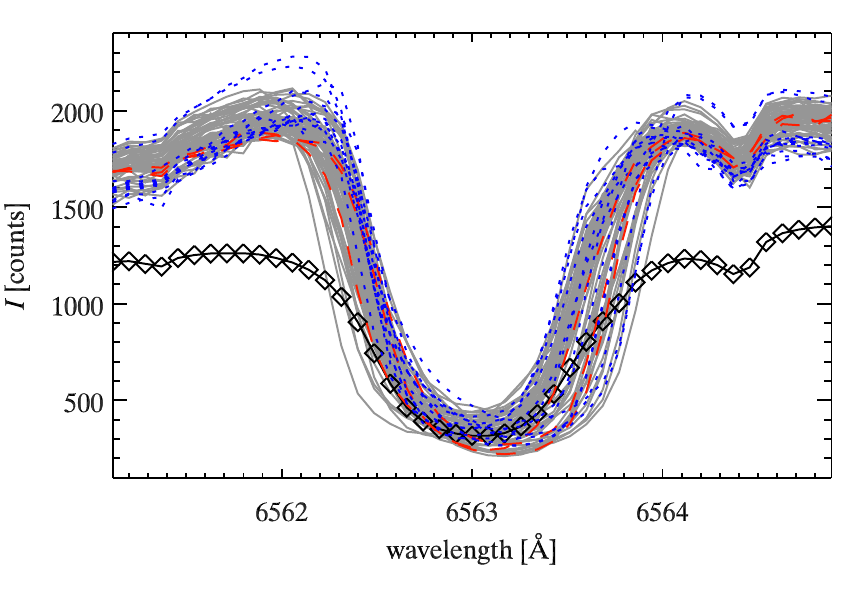}}
 \caption[]{\label{fig:ha-ca-profile}%
    Profiles for \EB\ detections in the \CaII\ blue wing
   (\emph{upper panel\/}), red wing (\emph{middle panel\/}) and in \Halpha\ (\emph{lower
   panel\/}).  Upper two panels: profiles of detections with blue-brighter-than-red
   \revv{(\emph{dashed blue\/})} or red-brighter-than-blue (\emph{dashed \revv{red}\/}) asymmetry by least
   10\% and all other detections (\emph{solid \revv{grey}\/}).
   Lower panel: profiles of detections with at least 5\% blue-brighter-than-red asymmetry
   (\emph{dotted \revv{blue}\/}), red-brighter-than-blue asymmetry (\emph{dashed \revv{red}\/}) and
   all other detections (\emph{solid grey\/}).
   In all panels the field-of-view average profile is also indicated ({\it solid black with diamonds\/}).
 }
\end{figure}

Figure~\ref{fig:ha-ca-profile} shows the detection-averaged profiles for \EBs\ observed in \revv{\CaII\ and
\Halpha\ along with the profile averaged over the full SST field-of-view for both lines.
The upper and middle panel show the profiles in the summed blue and red wings of \CaIR,
respectively.}
Most profiles appear to peak around $\pm$0.5--0.6\,\AA\ from line center and are asymmetric in shape. 
The blue wing detections have a general tendency to be brighter in the blue wing, while those in the
red wing show the opposite effect, although the picture is much more confusing in the latter case.
We also detect more \EBs\ for which the maximum brightness in the red wing is more than 10\%
larger than that in the blue wing (\rev{three} of which have a red wing brightness that exceeds that of the
blue wing by more than 20\%), than we do with the opposite asymmetry.
The lower panel of the same figure shows the detection-averaged profiles for \EBs\ in \Halpha\ in
the first data set.
Only 16 out of \rev{78} profiles show some sort of asymmetry, although less strongly than for the \CaIR\
detections (\ie\ all asymmetric profiles have the brightest wing exceeding the less bright wing by
no more than 10\%).
The majority of those asymmetric profiles have a blue-brighter-than-red wing signature.

\begin{figure*}[bht]
 \centerline{\includegraphics[width=\textwidth]{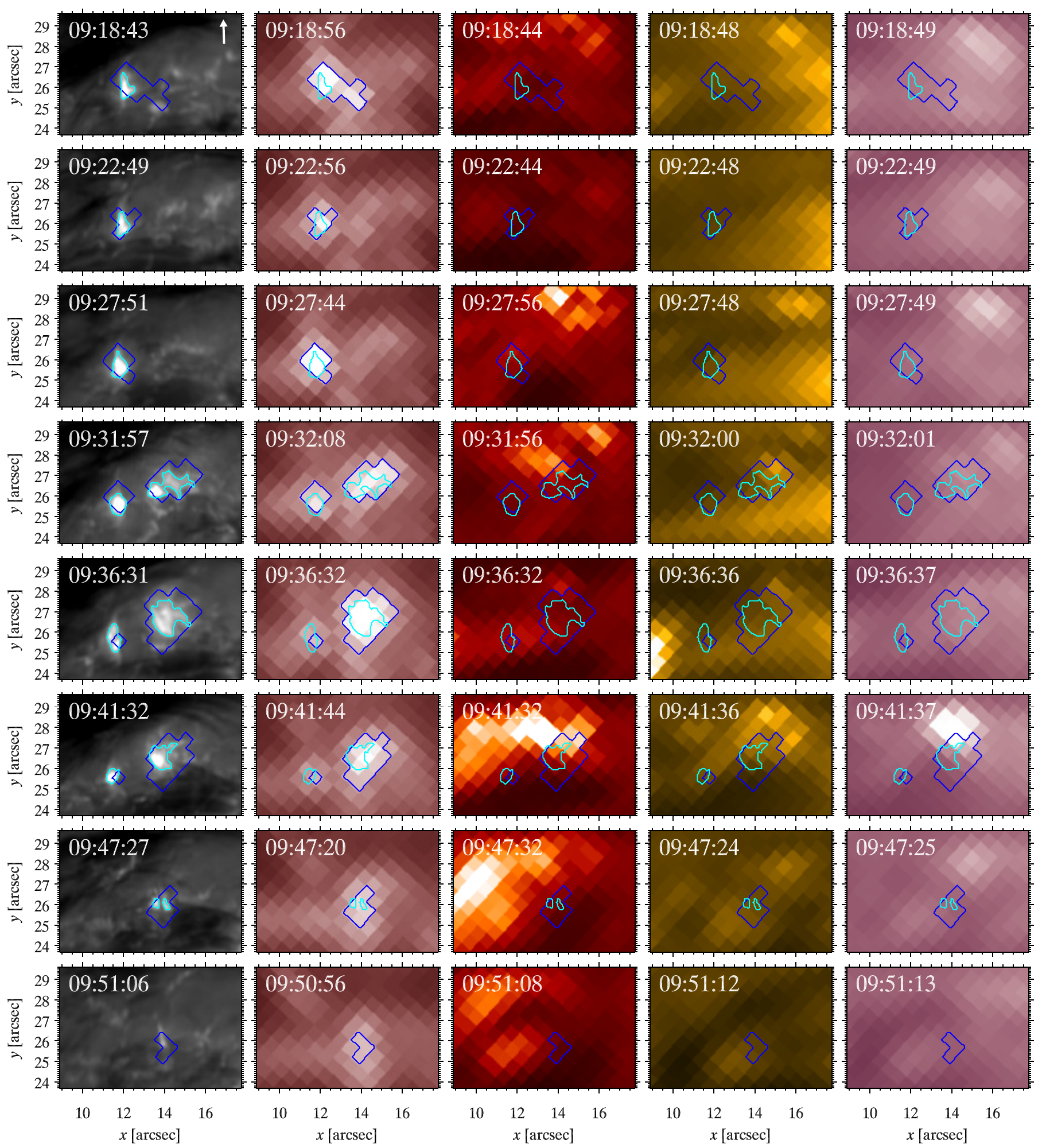}}
  \caption[]{\label{fig:sstsdo-timeseq}%
    Time evolution of a sequence of cutouts corresponding to region-of-interest 2.
    From left to right: CRISP images in the summed wings of \Halpha, AIA images in the continuum at
    1700\,\AA, \HeII, \FeIX, and \FeXIV.
    The cutouts are rotated counter-clockwise by 48\deg\ from their orientation in
    Fig.~\ref{fig:fov2} to obtain a near-vertical limbward direction,
    indicated by the white arrow in the top left panel.
    The contours specify results of the detection algorithm applied to \Halpha\ (\emph{azure\/})
    and 1700\,\AA\ (\emph{dark blue\/}).
    The time in UT is given in the upper left corner of the first column panels.}
\end{figure*}

\paragraph{Signature in AIA 1700\,\AA}
As already pointed out in the introduction, images taken in the
1700\,\AA\ continuum reproduce a similar patchwork of bright network
as observed in \CaII\ with localized brightenings that seem to
correspond closely to \EBs\ (cf.~Figs.~\ref{fig:fov1} and
\ref{fig:fov2}).
Figure~\ref{fig:sstsdo-timeseq} shows this in more detail by displaying part of the time evolution
of region-of-interest 2 in \Halpha\ and several AIA channels, with \Halpha\ and 1700\,\AA\ detection
contours overlaid (here we focus on the first two columns and postpone discussion of the remaining panels to
the end of this section). 
Comparison of the \Halpha\ and 1700\,\AA\ intensity images, as well as the detection contours on
both, shows that co-temporal brightenings can be found in 1700\,\AA, albeit at lower spatial
resolution than in the \Halpha\ data and, consequently, with differing morphology and extent.
However, a 5-$\sigma$ above average threshold does a relatively good job in recovering the brighter
\EBs\ as well as \EB\ conglomerates.

\begin{figure*}[bht]
 \centerline{
    \includegraphics[width=0.5\textwidth]{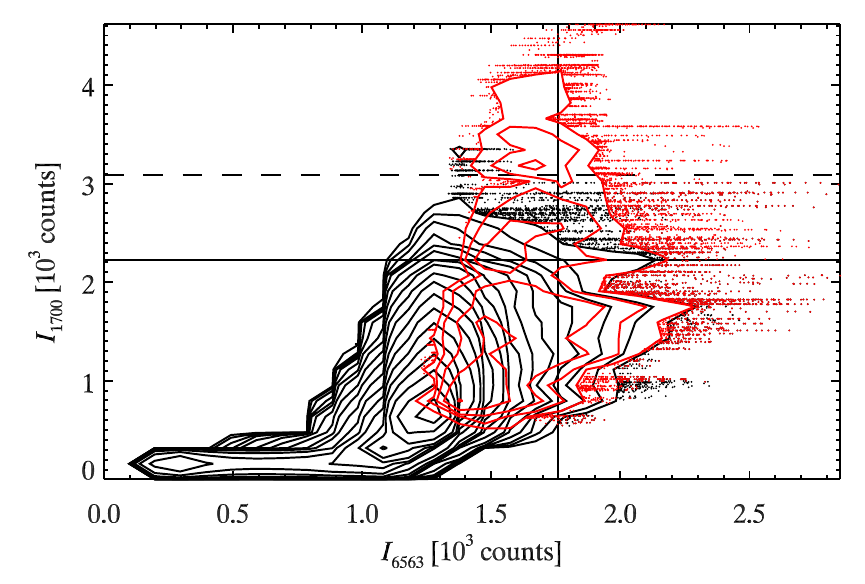}
    \includegraphics[width=0.5\textwidth]{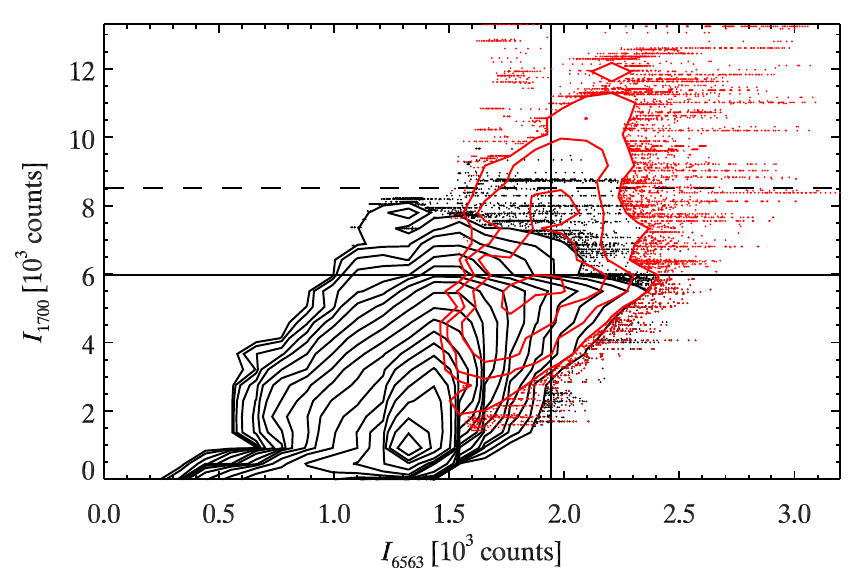}
 }
 \caption[]{\label{fig:sst-sdo-scatter}%
   CRISP \Halpha\ versus AIA 1700\,\AA\ scatter diagrams for data set
   1 (\emph{left panel\/}) and data set 2 (\emph{right panel\/}).
   As in Fig.~\ref{fig:ha-ca-scatter}, contours and points are drawn
   for all pixels (\emph{black\/}) and \Halpha\ detection pixels only
   (\emph{red\/}).
   The vertical solid line specifies the \rev{140\% of} average
   intensity threshold for \Halpha.
   The horizontal lines specify the 5-$\sigma$ (\emph{solid\/}) and 8-$\sigma$ (\emph{dashed\/})
   above average intensity for 1700\,\AA. 
 }
\end{figure*}

For a more detailed comparison we degraded our CRISP \Halpha\ data to
the much coarser pixel size of the AIA 1700\,\AA\ data, \ie\
0\farcs6\,px$^{-1}$.
Figure~\ref{fig:sst-sdo-scatter} shows scatter diagrams of these data,
where the red contours and points are based on the \Halpha\ detections
in the higher resolution CRISP data.
A further limiting criterion is that of the pixels exceeding
5-$\sigma$ above average intensity in 1700\,\AA, only those that
persist at such intensity for a period shorter than 5 minutes were
included.
Note that the number of pixels at SST resolution was retained in downsampling the \Halpha\ data and
that the 1700\,\AA\ were upscaled to the same number of pixels as the SST data (but retaining the
same SDO-sized pixel-shapes, \eg\ Fig.~\ref{fig:sstsdo-timeseq}).
The apparent quantization effect in the scatter clouds outside the contours is caused by both that
and by the fact that within the considered fields-of-view and time spans the high-intensity values
are not continuous.

We find a large degree of correlation for both data sets by excluding pixels with \Halpha\
brightness of less than about 1000\,counts, although not as tight as for the \Halpha-\CaIR\
comparison.
The low-1700\,\AA/lower-\Halpha\ intensity ``tongue'' in both panels
corresponds to the sunspots and is more pronounced for data set 1 due
to (1) the sunspot covering a larger portion of the field-of-view, and
(2) the absence of the lower-\Halpha/medium-1700\,\AA\ intensity bulge
in the lower left quadrant of the scatter diagram for data set 1.
The latter is probably a result of sampling brighter network at
1700\,\AA\ while strong absorptions are present at those pixels in the
summed wings of \Halpha\ for the second data set.

Considering only the pixels based on high-resolution \Halpha\
detections (red contours and points) we find that part of the \EBs\
would be recovered also in the lower resolution \Halpha\ data (even
more so in the second than in the first data set), but would have too
low 1700\,\AA\ brightness to distinguish them from regular network in
those data.
That the bulge of detection pixels falls below the \Halpha\ threshold
(\ie\ in the lower left quadrant) in both data sets is a result of
downsampling the CRISP data.
Interestingly though, many of the pixels above a 5-$\sigma$ threshold
in 1700\,\AA\ are the same pixels as recovered by the \Halpha\
detections and if increased to 8-$\sigma$ (corresponding to roughly
3100 and 8500\,counts for data sets 1 and 2, respectively) the overlap
would be near-perfect with respect to the brightest \Halpha\ \EBs.
Also, when applying the detection algorithm to the lower resolution \Halpha\ data, we recover
\rev{19} and \rev{23} detections (corresponding to \rev{about 24\% and 38\%} of the detections in the high-resolution
\Halpha\ data) in the first and second data set, respectively, comparable to the 1700\,\AA\
detection numbers.

\begin{figure*}[bhtp]
 \centerline{\includegraphics[width=\columnwidth]{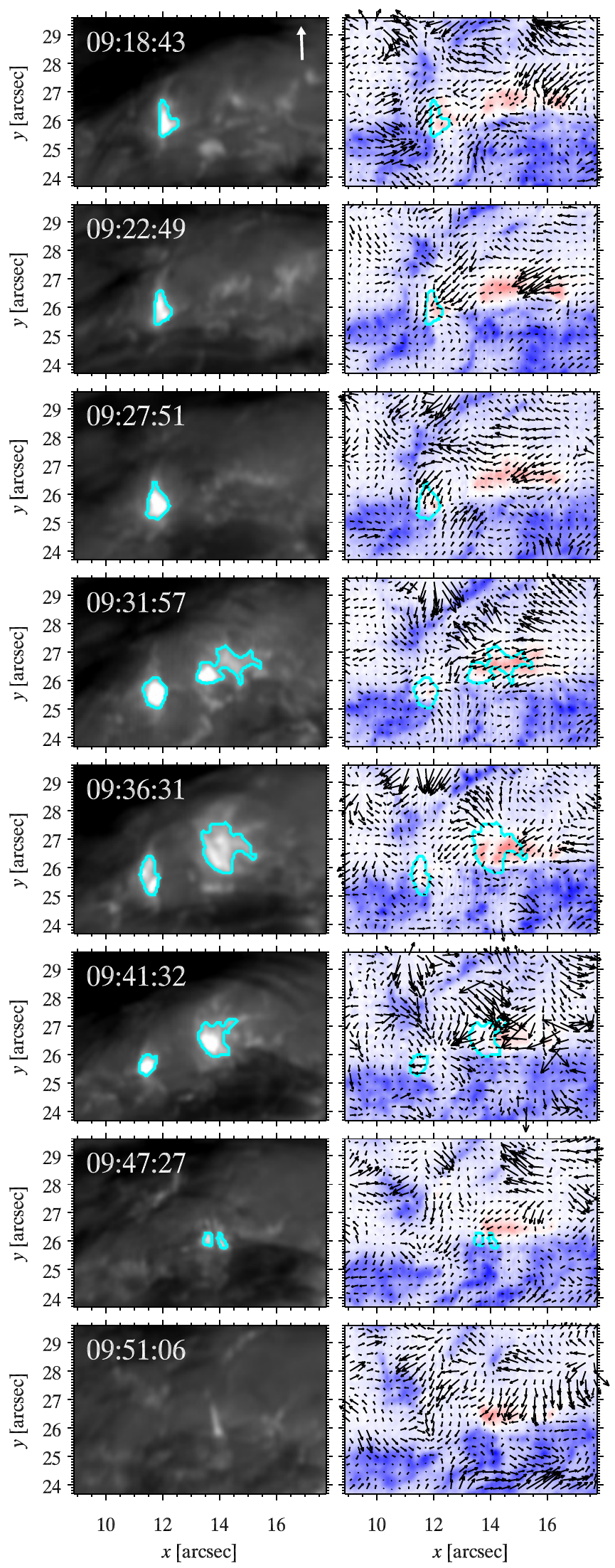}
   \includegraphics[width=\columnwidth]{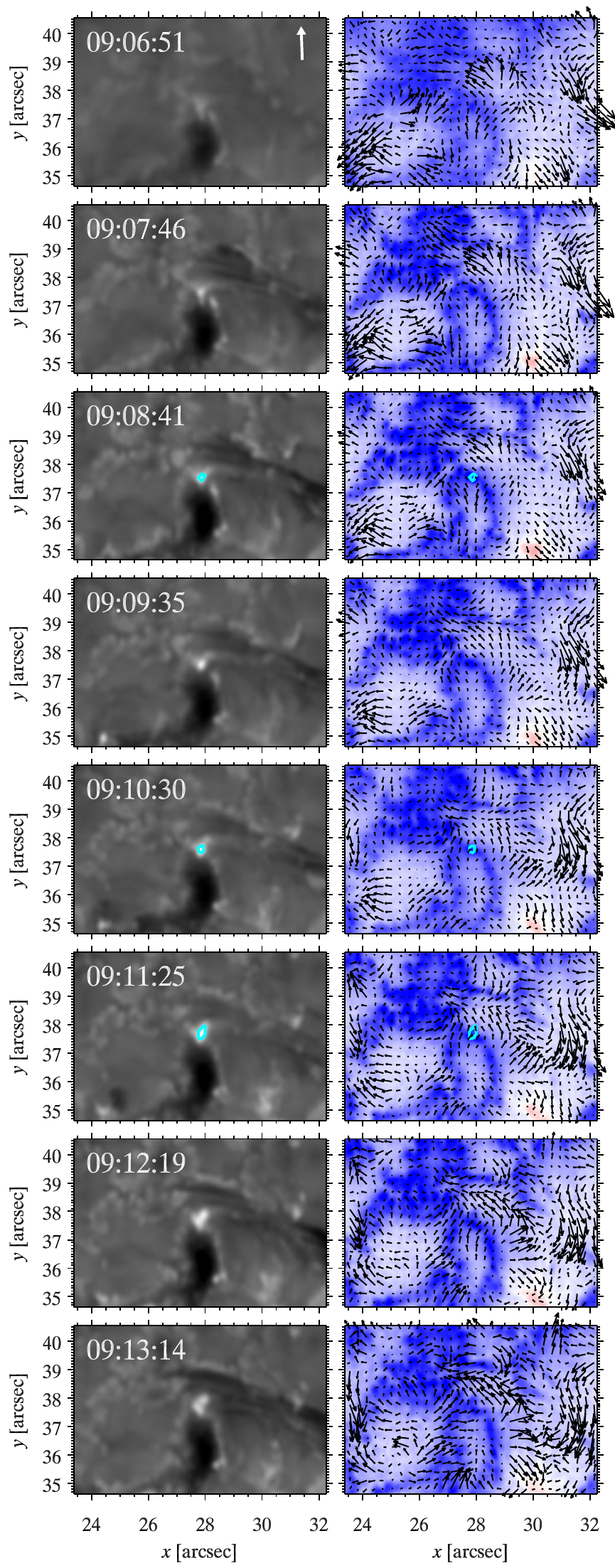}}
 \caption[]{\label{fig:sst-timeseq}%
   \EB\ evolution (azure contours) in region-of-interest 2
   (\emph{left-hand columns}) and region-of-interest 3
   (\emph{right-hand columns}).
   {\it First column\/}: \Halpha\ summed wings images for
   region-of-interest 2.
   {\it Second column\/}: \FeI\ Stokes-$V/I$ images for the same
   region of interest, with positive/negative values shown in red/blue
   and small black arrows indicating the surface flow field
   \rev{(we suggest zoom-in with a pdf viewer)}.
   {\it Third column\/}: \Halpha\ summed wings images for
   region-of-interest 3.
   {\it Fourth column\/}: \FeI\ Stokes-$V/I$ images for the same
   region of interest (format as for the second column panels).
   The Stokes-$V/I$ panels have been scaled to the full SST
   field-of-view at the first time step to enable comparison between
   the different regions-of-interest.
   The \Halpha\ panels have been scaled independently for each
   region-of-interest.
   The arrows in the top panel of the first and third columns indicate
   the limbward direction.
   The time in UT is given in the upper left corner of the first and
   third column panels.}
\end{figure*}

\begin{figure*}[t!]
 \centerline{\includegraphics[width=\columnwidth]{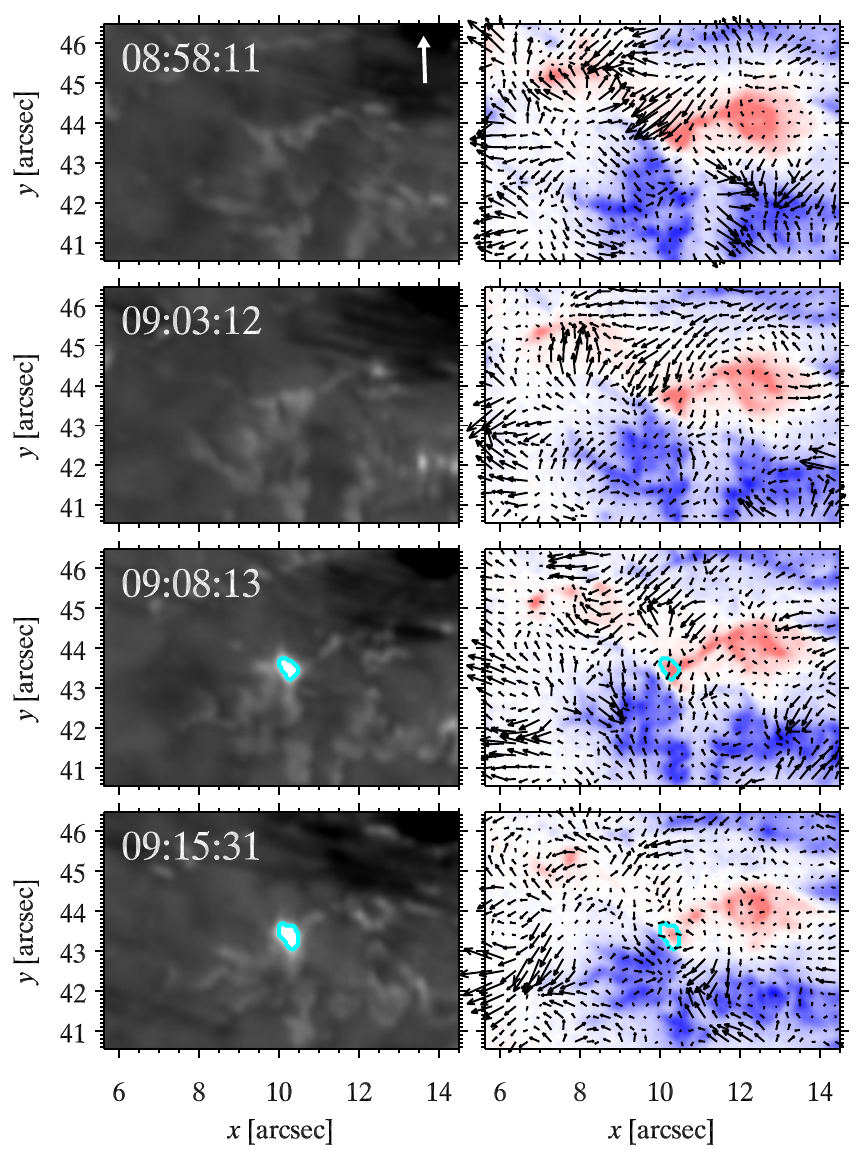}
   \includegraphics[width=\columnwidth]{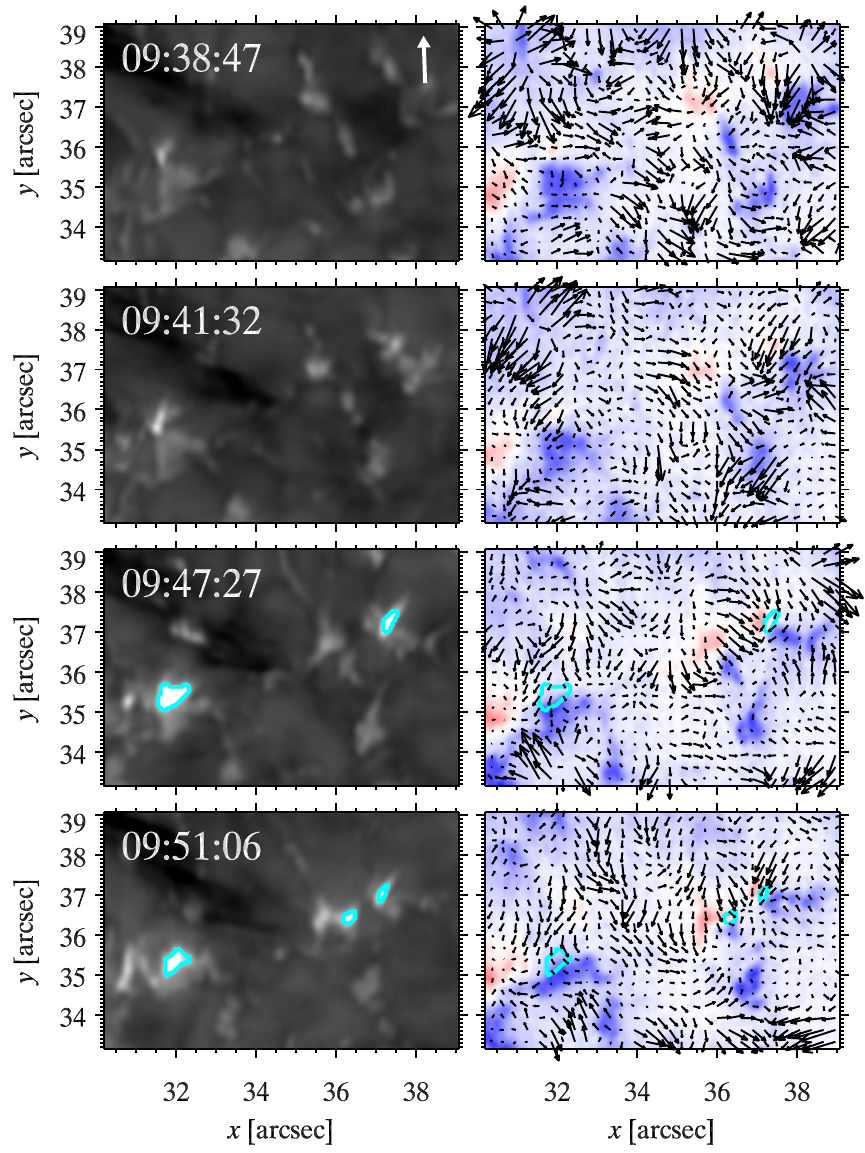}}
  \caption[]{\label{fig:sst-timeseq2}%
    \EB\ evolution in two more regions of interest marked in Fig.~\ref{fig:fov2}: region-of-interest
    4 ({\it left-hand columns\/}) and region-of-interest 5 ({\it right-hand columns\/}).
    Further format as for Fig.~\ref{fig:sst-timeseq}.
    }
\end{figure*}

\begin{figure}[t]
 \centerline{\includegraphics[width=\columnwidth]{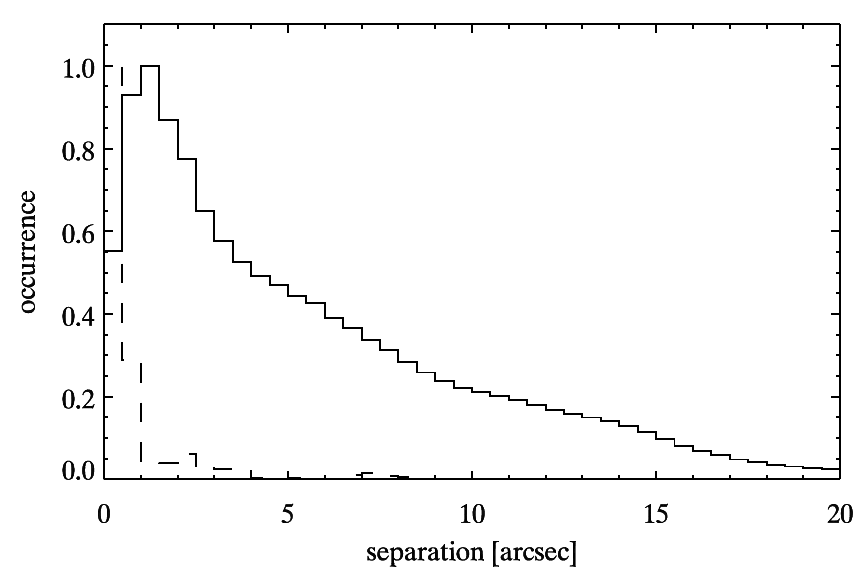}}
  \caption[]{\label{fig:disthist}%
    Distribution of the separation between opposite polarities for all pixels
    (\emph{solid line}) and \EB\ detection pixels (\emph{dashed line}) in the field of view of data
    set 2, excluding the sunspot, for frames with above average contrast.
    Both distributions have a bin size 0\farcs5 and have been scaled to their respective maximum
    values.
    }
\end{figure}

\paragraph{Occurrence location and magnetic field}
Figure~\ref{fig:sst-timeseq} shows the time sequence of a few \EBs\ in regions-of-interest 2 and 3.
Comparison of the \Halpha\ and \FeI\ Stokes-$V/I$ images shows that \EBs\ can generally be observed
at locations where opposite polarities meet, \ie\ the \EBs\ occur on the inversion
line between the opposite polarities and sometimes appear rooted in patches of both positive and
negative polarity.
Several examples of these properties are given in Figs.~\ref{fig:sst-timeseq} and \ref{fig:sst-timeseq2} for
regions-of-interest 2 through 5.
In particular, the \EB\ in the top three rows and the larger \EB\ cluster in the following three rows of
Fig.~\ref{fig:sst-timeseq} exhibit the rooting in opposite polarities, but it suggests
itself also for some of the \EBs\ in the lower panels of the two regions-of-interest in
Fig.~\ref{fig:sst-timeseq2}.
Quantitatively, this observation translates into Fig.~\ref{fig:disthist}, showing histograms of the
separation between opposite polarities for all pixels (solid line) and detection pixels only (dashed
line).
While the distribution of the opposite polarity separation for all pixels (\ie\ also including \EB\
detections), peaks in the \rev{1\farcs0--1\farcs5} bin and has an average of 5\farcs7, that of the \EB\
detection pixels alone is much narrower, peaking in the first bin and averaging at 0\farcs9.

In a small number of cases, an opposite polarity cannot be observed in the vicinity of the \EB, an
example of which is given in right-hand columns of Fig.~\ref{fig:sst-timeseq}.
The \EB\ in this figure seems to be rooted exclusively in a positive polarity patch with no sign of
any opposite polarity patch during (or prior to) its lifetime.
It is also notably smaller and less bright than the \EBs\ in regions-of-interest 2, 4 and 5, which
appears to be generally the case for \EBs\ in a (seemingly) unipolar magnetic field configuration.

Comparing the occurrence locations of \EBs\ to the surface flow field arrows in
Figs.~\ref{fig:sst-timeseq} and \ref{fig:sst-timeseq2} the \EBs\ seem to appear where the magnetic
field has been or is being pushed around.
Examples of this are numerous in said figures, \eg\ the flow field arrows
(1) above and behind the \EB\ in the top three rows of the left-hand columns in Fig.~\ref{fig:sst-timeseq},
(2) prior to and during the large \EB\ cluster in the same region-of-interest, starting at 09:31:57,
(3) nearby the \EB\ location in the first, third, sixth and seventh panels of the right-hand columns
in the same figure (although the flow field strength is noticeably smaller than in
other examples),
(4) above the negative polarity patch prior to the \EB\ in the left-hand columns of
Fig.~\ref{fig:sst-timeseq2},
(5) in the vicinity of the opposite polarity patches in the right-hand part of the right-hand
columns, both prior to and during the presence of an elongated \EB\ in the lower two rows,
(6) above the faint negative polarity patch (blue) prior to the \EB\ on the right-hand side of the same
region-of-interest.
Both the first and fifth are telltale examples of opposite polarities being pushed towards (and in the
latter case also alongside) each other, as the negative polarity patches move from
$(x,y)\!\approx\!(12.5,26.0)$ to $(x,y)\!\approx\!(11.8,25.5)$ in Fig.~\ref{fig:sst-timeseq} and from
$(x,y)\!\approx\!(37.5,38.0)$ to $(x,y)\!\approx\!(37.0,37.0)$ (towards and alongside a near-stationary
positive polarity) in Fig.~\ref{fig:sst-timeseq2}.
Moreover, the size and strength of the opposite polarity patches in the bipolar configurations are
greatly reduced during the \EB\ lifetimes and in some cases one of the polarities even vanishes
completely.  
Most notable examples of this are the first, second, fifth and sixth in the list above.

\paragraph{Comparison with outer atmosphere diagnostics}\label{sec:results_outer}
Considering that earlier studies have reported \EB-related surges, we investigated whether the
upper atmospheric AIA data showed any signs of perturbation by the underlying \EBs.
Figure~\ref{fig:sstsdo-timeseq} shows the time evolution of the second region-of-interest as
observed by both CRISP (in the summed wings of \Halpha) and AIA (in the continuum at 1700\,\AA,
\HeII, \FeIX, and \FeXIV).
This particular region-of-interest and time span (the same as Fig.~\ref{fig:sst-timeseq}) were
chosen as it shows both the brightest \EBs\ of the field-of-view during the time series and
presents the most tempting case for co-relating \EB\ presence with activity in the higher
atmosphere.

Comparison of the \Halpha\ and \HeII\ images generally shows no clear correspondence, even though
sometimes the \HeII\ images do display brightenings nearby, but not co-spatial with, the underlying
\EBs\ observed in the \Halpha\ wings (\eg\ the fourth, sixth and seventh rows in
Fig.~\ref{fig:sstsdo-timeseq}).
To a lesser extent this also holds for \FeIX\ and \FeXIV (cf.~the sixth and seventh rows).
However, when taking the bigger field-of-view shown in Fig.~\ref{fig:fov2} into account, the
aforementioned brightenings in \HeII, \FeIX, and \FeXIV\ seem rather a result of inflows along larger
scale structures and originate outside the field-of-view of the SST.
\EBs\ elsewhere in the field-of-view do not result in any perceivable signal in these diagnostics
either and running time-difference movies of the AIA data are equally inconclusive.

\section{Discussion}\label{sec:discussion}
\paragraph{Visibility in \Halpha}\label{sec:results_visibility}
\moved{ \EBs\ are traditionally defined as temporary brightenings of
  the outer \Halpha\ wings in emerging flux regions that have no
  signature in the \Halpha\ core
  (\citeads{1917ApJ....46..298E}). 
  However, the fourth and fifth panels of Fig.~\ref{fig:fov1} show
  many \Halpha-wing bright points near the spot that are not \EBs\ but
  mark more ubiquitous and stable kilogauss magnetic concentrations,
  similarly to G-band bright points
  (\citeads{2006A&A...452L..15L}; 
  \citeyearads{2006A&A...449.1209L}). 
  They are a subset of the magnetic bright points in the first three
  panels of Fig.~\ref{fig:fov1} where the first shows them the
  sharpest but with the lowest contrast \rev{(best with enlargement in
    a pdf viewer)}.
  The second panel shows them less sharp, due to higher-up radiation
  escape and scattering.
  The third panel renders them very similar to the second, except for
  the large difference in telescope resolution.
  The \Halpha\ wing panels in the second row show only those that are
  not shielded by overlying fibrils, with slight defocus caused by
  scattering in the transparent fibrils.
  Such ``pseudo Ellerman bombs'' are discussed in more detail in
  \citetads{2013JPhCS.440a2007R}. 
  Figures~\ref{fig:ha-ca-timeseq}, \ref{fig:sst-timeseq}, and \ref{fig:sst-timeseq2}
  demonstrate, as did Paper~I, that at
  the superb SST resolution in slanted limbward viewing, proper \EBs\
  show up as yet brighter, upright, short-lived, rapidly varying,
  elongated flames.
  We now prefer to define \EBs\ observationally as \Halpha-wing
  features with this flame morphology.
  Our detection algorithm is tailored to identify those features that
  we recognized visually as such flames.
}

\paragraph{Detection}
In this study, we define a fully automated detection algorithm to
select \EBs\ in both \Halpha\ and \CaII\ wing data.
This is a different approach from Paper~I, where only the network
bright points were detected automatically and \EBs\ selected manually
afterwards.
The detection constraints listed in Sect.~\ref{sec:detection} (kernel
brightness $>$\rev{155\% of} average, adjacent brightness
$>$\rev{140\% of} average, minimum size of 0\farcs2--0\farcs3, minimum
lifetime of $\sim$50\,s) are therefore different in several respects
from those presented in Paper~I, although most notably with respect to
the minimum lifetime (240\,s in Paper~I).
However, visual inspection of the results suggests that these
constraints are adequate in selecting events that, in addition to
excess wing brightness, display the telltale upright, flame-like
morphology \rev{while excluding  pseudo \EBs}.

The number of \EBs\ we find in our data is modest compared to some
earlier studies. 
For example, the birthrate of \citetads[][1.5\,min$^{-1}$ in an
18\arcsec$\times$24\arcsec\ region]{1987SoPh..108..227Z}
would predict about 510 and 590 \EBs\ in our first and second data set, respectively.
Even larger numbers were obtained in a recent study by
\citetads{2013SoPh..283..307N} 
who reported on the automated detection of 3570 \EBs\ in a 90 minute time series of
96\arcsec$\times$96\arcsec\ \Halpha\ data obtained with the IBIS instrument.
Their detection algorithm differs from ours mainly in its intensity threshold (130\% of
the average brightness), no minimum lifetime, and by considering events that fade below and brighten
again to above the threshold as separate events (\ie\ to account for seeing effects we allow up to
$\sim$50\,s of non-detection for spatially overlapping events to be considered the same).
The only way of detecting more events in our case would be to relax the intensity thresholds.
The \rev{140\% and 155\% of the field-of-view averaged brightness
  thresholds for both \Halpha\ and \CaIR\ that we used are} much more
restrictive than \rev{the threshold} adopted by
\citetads{2013SoPh..283..307N} 
or the \EB\ contrast range of 105\%--130\% reported earlier by
\citetads{2002ApJ...575..506G}. 
Conserving all other constraints, tests with lower than \rev{155\%}
initial thresholds
resulted in detection of many network bright points
and features that, in our opinion, are not \EBs.

\rev{ Most previous studies have reported \EB\ lifetimes between
  10--20\,min, in some cases even up to 30\,min (\eg\
  \citeads{2000ApJ...544L.157Q}; 
  \citeads{2006ApJ...643.1325F}; 
  \citeads{2008PASJ...60...95M}; 
  \citeads{2008ApJ...684..736W}; 
  \citeads{2011CEAB...35..181H}; 
  Paper~I).
  The average lifetime we find for \EB\ detections, 3.5--4\,min, is
  much shorter than that reported in Paper~I, which can largely be
  attributed to a more relaxed minimum lifetime threshold (240\,s in
  Paper~I versus $\sim$50\,s here), although a few seem to
  last as long as the earlier reported 20--30\,min.
  However, these longer detections typically display substructure and
  repetitive flaring and the lifetime of such substructure is much
  shorter than the ensemble lifetime.
  Similarly, the longer-lived detections in 1700\,\AA\ tend to coincide with multiple shorter-lived
  \Halpha\ detections, which may explain the longer average lifetime for 1700\,\AA\ detections.
  Our present results are comparable to those obtained for \EBs\ by
  \eg\ Pariat \etal\
  (\citeyearads{2007A&A...473..279P}, 
  reporting lifetimes between 1.5--7\,min with a peak around
  3--4\,min).
}

\rev{ The elongated substructures in our observations are typically
  1\arcsec\ tall and about 0\farcs2 wide.
  However, with 0.2--0.3\,arcsec$^{2}$, the average area of single
  \Halpha\ detections is much smaller than the 1--2\,arcsec$^{2}$
  reported before (\eg\
  \citeads{2002ApJ...575..506G}; 
  \citeads{2006ApJ...643.1325F}; 
  \citeads{2007A&A...473..279P}; 
  \citeads{2008PASJ...60..577M}). 
  The smaller sizes we find are most likely a result of the higher spatial
  resolution of the images, as single detections in AIA 1700\,\AA\
  data, with a pixel size of 0\farcs6\,px$^{-1}$, are similar in size
  to those reported earlier.  
}

\paragraph{Visibility in other diagnostics}
\citetads{2000ApJ...544L.157Q} 
and
\citetads{2002ApJ...575..506G} 
found that only about half of the \EBs\ identified in \Halpha\
correspond to brightenings in TRACE 1600\,\AA\ images, while
\citetads{2010MmSAI..81..646B} 
reported that all \EBs\ they found in \Halpha\ images from the
Dutch Open Telescope coincided with TRACE 1600\,\AA\
brightenings.
\citetads{2007A&A...473..279P}, 
comparing THEMIS \CaII\ and TRACE 1600\,\AA\ data, also found a good
correlation between the \EB\ locations in both diagnostics.

From our data we find that only part of the \EBs\ in \Halpha\ coincide with brightenings in
1600\,\AA\ and 1700\,\AA, in agreement with
\citetads{2000ApJ...544L.157Q} 
and
\citetads{2002ApJ...575..506G}. 
Although 1600\,\AA\ offers greater intensity contrast for the \EBs\ than 1700\,\AA, it suffers
noticeably from transition region contamination through the \CIV\ lines, complicating the
application of an automated detection algorithm and thereby rendering 1700\,\AA\ the better
AIA diagnostic for \EB\ detection.
Necessarily, only the larger and brighter \EBs\ or the enveloping haloes of multiple smaller \EBs\ 
are observable in the 1700\,\AA\ data (cf.~Figs.~\ref{fig:fov1}, \ref{fig:fov2}, and
\ref{fig:sstsdo-timeseq}), given that their spatial resolution is lower than that of the CRISP data.
This is further supported by the detection numbers and statistical evidence in
Fig.~\ref{fig:sst-sdo-scatter}, showing that most of the \Halpha\ detected \EBs\ cover pixels that
in 1700\,\AA\ have an intensity that cannot be differentiated from that of regular network and only
the brighter \Halpha\ pixels exceed the 5-$\sigma$ threshold in 1700\,\AA\ (more so for the second
than for the first data set).
Notwithstanding, our results suggest that modifying the detection algorithm from
Sect.~\ref{sec:detection} to incorporate a brightness threshold of 8-$\sigma$ above average, as well
as an upper limit of 5 minutes on the lifetime would provide an effective recipe to detect \EBs\
in 1700\,\AA, possibly expanded with a maximum size to prevent detection of flare-like events.
Even though not all \EBs\ visible in the CRISP \Halpha\ data would be recovered this way, AIA
1700\,\AA\ has the clear advantage of providing near-continuous imaging of the entire Earth-side
solar disk.

Comparing the detections in \Halpha\ and \CaII\ shows that only part of the detections
in the former have a good spatial overlap with those in the \CaIR\ blue and red wing.
\revv{Such discrepancy in morphology between \EBs\ in  these two lines has not been reported
before.}
For most \EBs\ we find a clear morphological dissimilarity (cf.~Figs.~\ref{fig:ha-ca-timeseq} and
\ref{fig:ha-ca-scatter}).
This is the case both when comparing \Halpha\ with either wing of \CaIR, as well as the
wings of the latter with each other.  
As illustrated in Fig.~\ref{fig:ha-ca-timeseq}, in some cases the brightenings in \CaIR\ lag behind
those in \Halpha\ (with respect to the proper motion of the \EBs) or appear on top of those.
We speculate this may be due to projection effects, as we find that the well-overlapping detections
concern \EBs\ that have a proper motion roughly along the line-of-sight, whereas those that overlap
only partly detect \EBs\ moving at an angle to the line-of-sight.
\revv{Additional effects explaining the spatial differences could be the difference in recombination
rate and opacity between the two lines.
Both \Halpha\ and \CaII\ probably show the afterglow of subsequent recombination
\citepads{2013JPhCS.440a2007R}, 
while the reconnection likely takes place on very small spatial and temporal scales.
Different recombination rates and \CaIR\ having opacity surrounding the \EB\ where \Halpha\ has
none, may cause significant differences in the morphology of \EBs\  as observed in both lines.
}

Some of the \EBs\ detected in the \CaIR\ blue wing are not detected in
the red wing, and vice versa, which may be explained by an asymmetry
in their respective spectral profiles.
Such asymmetry has been known for a long time from \EB\ studies in
\Halpha\ (\eg\ \citeads{1968mmsf.conf...71S}; 
\citeads{1968mmsf.conf..109E}; 
\citeads{1970SoPh...11..276K}; 
\citeads{1972SoPh...26...94B}; 
\citeads{1983SoPh...87..135K}), 
\CaII\ (\eg\
\citeads{2006ApJ...643.1325F}; 
\citeads{2006SoPh..235...75S}), 
and \CaIIH\
(\citeads{2010PASJ...62..879H}). 
The blue-asymmetry (\ie\ the blue wing brighter than the red wing) is
the most common, but opposite asymmetries (or lack of a strong trend)
have also been reported
(\citeads{2006ApJ...643.1325F}; 
\citeads{2006SoPh..235...75S}; 
\citeads{2007A&A...473..279P}), 
as well as asymmetry changes within \EBs\ during their lifetimes
\citepads{2010PASJ...62..879H}. 

We find no strong evidence of such asymmetry in the detection-averaged
\Halpha\ profiles of data set 1.
Only about 20\% of the profiles have an appreciable wing-excess
asymmetry (the majority of those are blue-asymmetric, in accordance
with previous reports), but none have the intensity in one wing
exceeding that in the other by more than 10\%.
However, in contrast to the findings of
\citetads{2006ApJ...643.1325F} 
and \citetads{2007A&A...473..279P}, 
the \EBs\ in our \CaIR\ images appear to suffer more from these
asymmetries, being both more prevalent and stronger (cf.~the top
panels of Fig.~\ref{fig:ha-ca-profile}).
It should be noted that the \CaII\ wing images we present in this work
have been obtained by summing over a small range around $\pm$0.6\,\AA\
from \CaIR\ line center, which is further out than the $\pm$0.35\,\AA\
where \citetads{2007A&A...473..279P} 
reported intensity peaks in the \CaIR\ spectrum (while the spectral
passband is comparable).
Although some of the \CaIR\ brightenings we observe in the blue and
red wings appear to be visible closer to line center as well, the view
at $\pm$0.35\,\AA\ is permeated with fibrillar structures comparable
to \Halpha$\pm$0.5\,\AA\ (panels 5 and 6 in Fig.~\ref{fig:fov2}),
complicating the clear identification of \EBs\ (Paper~I;
\citeads{2013JPhCS.440a2007R}). 

The profile asymmetry is well explained as a result of overlying,
Doppershifted fibrils (as pointed out in
\citeads{1972SoPh...26...94B}, 
\citeads{1983SoPh...87..135K}, 
\citeads{1997A&A...322..653D}, 
Paper~I, and 
\citeads{2013JPhCS.440a2007R}). 
The superpenumbral fibrils on the disk-center side of the sunspot show
a stronger absorption in the red wing, while those on the limb-side
are darker in the blue wing, \ie\ the line core is shifted blue-wards
on the limb-side and red-wards on the disk-center side
(signature of the inverse Evershed flow along those
fibrils, \citeads{1909Obs....32..291E}). 
The fifth and sixth panels of Fig.~\ref{fig:fov1} illustrate this clearly.
\revv{As the atomic mass of calcium is larger than that of hydrogen, the thermal width of \CaII\ is
much smaller and its sensitivity to this effect is consequently larger.
The asymmetries that result by the overlying Dopplershifted fibrils eating up the emission signal is
thus more pronounced in \CaIR\ and may also explain} 
why less \EBs\ were detected in in \CaIR.

Full explanation of the different appearance of \EBs\ in different
diagnostics requires detailed radiative transfer modeling 
\revv{while such morphological differences may well provide important
  constraints to numerical \EB\ simulation in the first place.
  However, such studies are} beyond the scope of this paper.
Some of the line formation suggestions of
\citetads{2013JPhCS.440a2007R} 
are presently being tested by the Oslo group. 

\paragraph{Triggering}
Magnetic reconnection has been proposed in many previous studies as the driving mechanism of \EBs,
although the actual field topology is still debated
(\eg\ 
\citeads{2002ApJ...575..506G}; 
\citeads{2004ApJ...614.1099P}; 
\citeads{2008ApJ...684..736W}; 
\citeads{2008PASJ...60..577M}; 
\citeads{2010PASJ...62..879H}; 
\citeads{2012EAS....55..115P}). 
Evidence for bidirectional flows in \EBs\  has been found in \CaIIH\ 
(\citeads{2008PASJ...60...95M}) 
and \Halpha\ data (Paper~I).
Combined with the jet-like structure reported here and in Paper~I, this could be indicative of
reconnection by a mechanism similar to that in so-called anemone jets
(\citeads{2007Sci...318.1591S}; 
\citeads{2010PASJ...62..901M}; 
\citeads{2011ApJ...731...43N}). 
The majority of the \EBs\ are found along magnetic polarity inversion lines
(\eg\
\citeads{2006ApJ...643.1325F}; 
\citeads{2007A&A...473..279P}; 
\citeads{2008PASJ...60..577M}; 
\citeads{2010PASJ...62..879H}), 
but an appreciable fraction is observed in apparently unipolar configurations
(\citeads{2000ApJ...544L.157Q}; 
\citeads{2002ApJ...575..506G}; 
\citeads{2008ApJ...684..736W}). 

We find that most \EBs\ in the field-of-view of the second data set occur where opposite polarities
meet, although not all locations with adjacent bipolar fields result in an \EB. 
In agreement with 
\citetads{2010PASJ...62..879H}, 
we observe that one or both of the polarity patches decreases in strength during the \EB\ lifetime.
Typically the smaller patch also shrinks, sometimes to the point that it completely
disappears, but this could be a detection sensitivity effect of our \FeI\ Stokes-$V/I$
data.
In addition, our simultaneous photospheric surface flow maps show that patches of opposite polarity
are in many events driven towards each other, either head-on or in a shearing motion.
This is consistent with a configuration similar to that in cartoon 1 in Fig.~17 of 
\citeads{2008ApJ...684..736W} 
or Fig.~19 in 
\citetads{2010PASJ...62..879H}, 
although it does not rule out a scenario in which flux emerges resistively and reconnects with
existing fields
(\eg\
\citeads{2007ApJ...657L..53I}; 
\citeads{2012EAS....55..115P}). 
The surface flows are typically strongest just prior to the detection of the \EBs\ with the \EBs\
moving in the direction of the flows.
The latter was already described in Paper~I, but here we provide further quantitative evidence.

However, not all \EB\ detections correspond to locations with clear
opposite polarity patches and we find a number of them in apparently
unipolar regions, close to the sunspot penumbra (\eg\ the event in the
third and fourth columns of Fig.~\ref{fig:sst-timeseq}).
This could be indicative of unipolar shearing reconnection (\eg\
\citeads{2008ApJ...684..736W}) 
or, alternatively, the opposite polarity is too weak to be detected.
The latter is a realistic possibility, considering the typically
smaller size and lower brightness of unipolar events, as well as that
they tend to occur close to the penumbra, where the field is stronger.

\paragraph{Effect on the upper atmosphere}
\rev{As noted in the introduction, correspondence of \EBs\ with surges
  and high-energy events in the upper atmosphere has been reported
  but seems not ubiquitous 
  (\citeads{2004ApJ...601..530S}). 
  In Paper~I only 2 out of 17 \EBs\ presented a possibly related
  surge; here we found none.
} 
In our comparison of the \rev{high-cadence upper}
atmospheric AIA data in \HeII, \FeIX, and \FeXIV\ with the \EB\
locations in \Halpha\ we find no conclusive evidence \rev{for
  co-related upper atmosphere signals}. 
The most tempting case was found in data set 2, where multiple
\EBs\ are occurring while co-temporal and nearly co-spatial
brightenings are observed in \HeII\ and \FeXIV, and to some extent in
\FeIX\ (cf.~Fig.~\ref{fig:sstsdo-timeseq}).
Although these brightenings could be linked to the underlying \EBs,
the dynamics in the larger field-of-view suggest they are rather
related to flows along the larger loop-like structures that extend to
beyond the field-of-view of the SST.
Also, equally bright \EBs\ elsewhere in the field-of-view produce no
perceivable effects in either 304\,\AA, 171\,\AA\ or 211\,\AA.
Similarly, \Halpha\ and \CaII\ images closer to line center, \ie\
sampling greater heights than the far wings, show no evidence for
\EB-related surges. 
Hence, our data offer no support for the earlier claimed connections
of \EBs\ to microflares, flaring arch filaments, or surges.

\section{Conclusions}  \label{sec:conclusion}
In this paper we have studied two active regions using high-resolution
CRISP imaging spectroscopy in \Halpha\ and \CaII, imaging
spectropolarimetry in \FeI, and AIA imaging in the UV-continua at
1600\,\AA\ and 1700\,\AA, in \HeII, \FeIX, and \FeXIV.
The co-spatial and co-temporal \FeI\ Stokes-$V/I$ data have allowed us
to expand on the work previously presented in Paper~I, by considering
the relation of \EBs\ to the underlying magnetic field in more depth.
On the other hand, simultaneous \CaII\ and AIA 1700\,\AA\ imaging has
provided a multi-diagnostic view of \EBs, while the AIA 304\,\AA,
171\,\AA, and 211\,\AA\ data have enabled us to study possible \EB\
effects on the upper atmosphere.

We find that a clear majority of the \EBs\ occurs where positive and
negative polarities are driven together by the photospheric surface
flows, enabling opposite polarity cancelation. 
A small number is also found in unipolar regions where shearing
reconnection may take place.
In either case, these results strengthen the case for a scenario in
which frozen-in fields are carried by the moat flow around sunspots,
causing \EBs\ as they reconnect.
Morphologically, \EBs\ often appear dissimilar in \CaII\ and \Halpha,
and we detect far fewer \EBs\ in \CaIR\ than in \Halpha.
Both may be due to the larger sensitivity of \CaIR\ to Dopplershifts
of the superpenumbral fibrils overhead, consequently producing the
strong asymmetric \EB\ profiles.
The brighter \EBs\ also have distinguishing signature in AIA
1700\,\AA, although none of the finer substructure is observed in the
lower resolution AIA images.
However, even though automated detections in AIA 1700\,\AA\ would miss
out on two thirds to three quarters of the \EBs\ visible in the
\Halpha\ wings, it may offer a straightforward way to track flux
emergence in large active regions or even full-disk images, as well as
enable the build-up of long-term, full-disk \EB-statistics.
Finally, no convincing evidence is found for influence from underlying
\EBs\ on the outer atmosphere and we therefore conclude that \EBs\ are
purely photospheric phenomena, incapable of breaking through the
overlying canopy.

\acknowledgments 
${}$\newline

We thank Jaime de la Cruz Rodr\' iguez and Eamon Scullion for their
help during the observations.
We also thank Eamon Scullion for providing the procedures for and
helping with the co-alignment of the SDO/AIA and SST/CRISP data.
We made much use of NASA's Astrophysics Data System Bibliographic
Services.
The Swedish 1-m Solar Telescope is operated on the island of
La Palma by the Institute for Solar Physics of Stockholm
University in the Spanish Observatorio del Roque de los
Muchachos of the Instituto de Astrof\' isica de Canarias.

\bibliographystyle{apj}
 \bibliography{eb2} 

\begin{thebibliography}{60}
\expandafter\ifx\csname natexlab\endcsname\relax\def\natexlab#1{#1}\fi

\bibitem[{{Archontis} \& {Hood}(2009)}]{2009A&A...508.1469A}
{Archontis}, V., \& {Hood}, A.~W. 2009, \aap, 508, 1469

\bibitem[{{Berlicki} {et~al.}(2010){Berlicki}, {Heinzel}, \&
  {Avrett}}]{2010MmSAI..81..646B}
{Berlicki}, A., {Heinzel}, P., \& {Avrett}, E.~H. 2010, \memsai, 81, 646

\bibitem[{{Bernasconi} {et~al.}(2002){Bernasconi}, {Rust}, {Georgoulis}, \&
  {Labonte}}]{2002SoPh..209..119B}
{Bernasconi}, P.~N., {Rust}, D.~M., {Georgoulis}, M.~K., \& {Labonte}, B.~J.
  2002, \solphys, 209, 119

\bibitem[{{Bruzek}(1972)}]{1972SoPh...26...94B}
{Bruzek}, A. 1972, \solphys, 26, 94

\bibitem[{{Cheung} {et~al.}(2008){Cheung}, {Sch{\"u}ssler}, {Tarbell}, \&
  {Title}}]{2008ApJ...687.1373C}
{Cheung}, M.~C.~M., {Sch{\"u}ssler}, M., {Tarbell}, T.~D., \& {Title}, A.~M.
  2008, \apj, 687, 1373

\bibitem[{{Dara} {et~al.}(1997){Dara}, {Alissandrakis}, {Zachariadis}, \&
  {Georgakilas}}]{1997A&A...322..653D}
{Dara}, H.~C., {Alissandrakis}, C.~E., {Zachariadis}, T.~G., \& {Georgakilas},
  A.~A. 1997, \aap, 322, 653

\bibitem[{{de la Cruz Rodr{\'{\i}}guez}(2012)}]{2012PhDT.........8D}
{de la Cruz Rodr{\'{\i}}guez}, J. 2012, PhD thesis, PhD Thesis, 2012

\bibitem[{{Ellerman}(1917)}]{1917ApJ....46..298E}
{Ellerman}, F. 1917, \apj, 46, 298

\bibitem[{{Engvold} \& {Maltby}(1968)}]{1968mmsf.conf..109E}
{Engvold}, O., \& {Maltby}, P. 1968, in Mass Motions in Solar Flares and
  Related Phenomena, ed. Y.~{Oehman}, 109

\bibitem[{{Evershed}(1909)}]{1909Obs....32..291E}
{Evershed}, J. 1909, The Observatory, 32, 291

\bibitem[{{Fang} {et~al.}(2006){Fang}, {Tang}, {Xu}, {Ding}, \&
  {Chen}}]{2006ApJ...643.1325F}
{Fang}, C., {Tang}, Y.~H., {Xu}, Z., {Ding}, M.~D., \& {Chen}, P.~F. 2006,
  \apj, 643, 1325

\bibitem[{{Georgoulis} {et~al.}(2002){Georgoulis}, {Rust}, {Bernasconi}, \&
  {Schmieder}}]{2002ApJ...575..506G}
{Georgoulis}, M.~K., {Rust}, D.~M., {Bernasconi}, P.~N., \& {Schmieder}, B.
  2002, \apj, 575, 506

\bibitem[{{Guglielmino} {et~al.}(2010){Guglielmino}, {Bellot Rubio},
  {Zuccarello}, {Aulanier}, {Vargas Dom{\'{\i}}nguez}, \&
  {Kamio}}]{2010ApJ...724.1083G}
{Guglielmino}, S.~L., {Bellot Rubio}, L.~R., {Zuccarello}, F., {et~al.} 2010,
  \apj, 724, 1083

\bibitem[{{Hashimoto} {et~al.}(2010){Hashimoto}, {Kitai}, {Ichimoto}, {Ueno},
  {Nagata}, {Ishii}, {Hagino}, {Komori}, {Nishida}, {Matsumoto}, {Otsuji},
  {Nakamura}, {Kawate}, {Watanabe}, \& {Shibata}}]{2010PASJ...62..879H}
{Hashimoto}, Y., {Kitai}, R., {Ichimoto}, K., {et~al.} 2010, \pasj, 62, 879

\bibitem[{{Henriques}(2012)}]{2012A&A...548A.114H}
{Henriques}, V.~M.~J. 2012, \aap, 548, A114

\bibitem[{{Herlender} \& {Berlicki}(2011)}]{2011CEAB...35..181H}
{Herlender}, M., \& {Berlicki}, A. 2011, Central European Astrophysical
  Bulletin, 35, 181

\bibitem[{{Isobe} {et~al.}(2007){Isobe}, {Tripathi}, \&
  {Archontis}}]{2007ApJ...657L..53I}
{Isobe}, H., {Tripathi}, D., \& {Archontis}, V. 2007, \apjl, 657, L53

\bibitem[{{Kitai}(1983)}]{1983SoPh...87..135K}
{Kitai}, R. 1983, \solphys, 87, 135

\bibitem[{{Koval} \& {Severny}(1970)}]{1970SoPh...11..276K}
{Koval}, A.~N., \& {Severny}, A.~B. 1970, \solphys, 11, 276

\bibitem[{{Leenaarts} {et~al.}(2006{\natexlab{a}}){Leenaarts}, {Rutten},
  {Carlsson}, \& {Uitenbroek}}]{2006A&A...452L..15L}
{Leenaarts}, J., {Rutten}, R.~J., {Carlsson}, M., \& {Uitenbroek}, H.
  2006{\natexlab{a}}, \aap, 452, L15

\bibitem[{{Leenaarts} {et~al.}(2006{\natexlab{b}}){Leenaarts}, {Rutten},
  {S{\"u}tterlin}, {Carlsson}, \& {Uitenbroek}}]{2006A&A...449.1209L}
{Leenaarts}, J., {Rutten}, R.~J., {S{\"u}tterlin}, P., {Carlsson}, M., \&
  {Uitenbroek}, H. 2006{\natexlab{b}}, \aap, 449, 1209

\bibitem[{{Lemen} {et~al.}(2012){Lemen}, {Title}, {Akin}, {Boerner}, {Chou},
  {Drake}, {Duncan}, {Edwards}, {Friedlaender}, {Heyman}, {Hurlburt}, {Katz},
  {Kushner}, {Levay}, {Lindgren}, {Mathur}, {McFeaters}, {Mitchell}, {Rehse},
  {Schrijver}, {Springer}, {Stern}, {Tarbell}, {Wuelser}, {Wolfson}, {Yanari},
  {Bookbinder}, {Cheimets}, {Caldwell}, {Deluca}, {Gates}, {Golub}, {Park},
  {Podgorski}, {Bush}, {Scherrer}, {Gummin}, {Smith}, {Auker}, {Jerram},
  {Pool}, {Soufli}, {Windt}, {Beardsley}, {Clapp}, {Lang}, \&
  {Waltham}}]{2012SoPh..275...17L}
{Lemen}, J.~R., {Title}, A.~M., {Akin}, D.~J., {et~al.} 2012, \solphys, 275, 17

\bibitem[{{Lites}(1987)}]{1987ApOpt..26.3838L}
{Lites}, B.~W. 1987, \ao, 26, 3838

\bibitem[{{Litvinenko}(1999)}]{1999ApJ...515..435L}
{Litvinenko}, Y.~E. 1999, \apj, 515, 435

\bibitem[{{Madjarska} {et~al.}(2009){Madjarska}, {Doyle}, \& {De
  Pontieu}}]{2009ApJ...701..253M}
{Madjarska}, M.~S., {Doyle}, J.~G., \& {De Pontieu}, B. 2009, \apj, 701, 253

\bibitem[{{Matsumoto} {et~al.}(2008{\natexlab{a}}){Matsumoto}, {Kitai},
  {Shibata}, {Otsuji}, {Naruse}, {Shiota}, \& {Takasaki}}]{2008PASJ...60...95M}
{Matsumoto}, T., {Kitai}, R., {Shibata}, K., {et~al.} 2008{\natexlab{a}},
  \pasj, 60, 95

\bibitem[{{Matsumoto} {et~al.}(2008{\natexlab{b}}){Matsumoto}, {Kitai},
  {Shibata}, {Nagata}, {Otsuji}, {Nakamura}, {Watanabe}, {Tsuneta}, {Suematsu},
  {Ichimoto}, {Shimizu}, {Katsukawa}, {Tarbell}, {Lites}, {Shine}, \&
  {Title}}]{2008PASJ...60..577M}
---. 2008{\natexlab{b}}, \pasj, 60, 577

\bibitem[{{Morita} {et~al.}(2010){Morita}, {Shibata}, {Ueno}, {Ichimoto},
  {Kitai}, \& {Otsuji}}]{2010PASJ...62..901M}
{Morita}, S., {Shibata}, K., {Ueno}, S., {et~al.} 2010, \pasj, 62, 901

\bibitem[{{Nelson} {et~al.}(2013){Nelson}, {Doyle}, {Erd{\'e}lyi}, {Huang},
  {Madjarska}, {Mathioudakis}, {Mumford}, \& {Reardon}}]{2013SoPh..283..307N}
{Nelson}, C.~J., {Doyle}, J.~G., {Erd{\'e}lyi}, R., {et~al.} 2013, \solphys,
  283, 307

\bibitem[{{Nishizuka} {et~al.}(2011){Nishizuka}, {Nakamura}, {Kawate}, {Singh},
  \& {Shibata}}]{2011ApJ...731...43N}
{Nishizuka}, N., {Nakamura}, T., {Kawate}, T., {Singh}, K.~A.~P., \& {Shibata},
  K. 2011, \apj, 731, 43

\bibitem[{{Nozawa} {et~al.}(1992){Nozawa}, {Shibata}, {Matsumoto}, {Sterling},
  {Tajima}, {Uchida}, {Ferrari}, \& {Rosner}}]{1992ApJS...78..267N}
{Nozawa}, S., {Shibata}, K., {Matsumoto}, R., {et~al.} 1992, \apjs, 78, 267

\bibitem[{{Pariat} {et~al.}(2004){Pariat}, {Aulanier}, {Schmieder},
  {Georgoulis}, {Rust}, \& {Bernasconi}}]{2004ApJ...614.1099P}
{Pariat}, E., {Aulanier}, G., {Schmieder}, B., {et~al.} 2004, \apj, 614, 1099

\bibitem[{{Pariat} {et~al.}(2006){Pariat}, {Aulanier}, {Schmieder},
  {Georgoulis}, {Rust}, \& {Bernasconi}}]{2006AdSpR..38..902P}
---. 2006, Advances in Space Research, 38, 902

\bibitem[{{Pariat} {et~al.}(2012{\natexlab{a}}){Pariat}, {Masson}, \&
  {Aulanier}}]{2012ASPC..455..177P}
{Pariat}, E., {Masson}, S., \& {Aulanier}, G. 2012{\natexlab{a}}, in
  Astronomical Society of the Pacific Conference Series, Vol. 455, 4th Hinode
  Science Meeting: Unsolved Problems and Recent Insights, ed. L.~{Bellot
  Rubio}, F.~{Reale}, \& M.~{Carlsson}, 177

\bibitem[{{Pariat} {et~al.}(2007{\natexlab{a}}){Pariat}, {Schmieder},
  {Berlicki}, {Deng}, {Mein}, {L{\'o}pez Ariste}, \&
  {Wang}}]{2007A&A...473..279P}
{Pariat}, E., {Schmieder}, B., {Berlicki}, A., {et~al.} 2007{\natexlab{a}},
  \aap, 473, 279

\bibitem[{{Pariat} {et~al.}(2007{\natexlab{b}}){Pariat}, {Schmieder},
  {Berlicki}, \& {L{\'o}pez Ariste}}]{2007ASPC..368..253P}
{Pariat}, E., {Schmieder}, B., {Berlicki}, A., \& {L{\'o}pez Ariste}, A.
  2007{\natexlab{b}}, in Astronomical Society of the Pacific Conference Series,
  Vol. 368, The Physics of Chromospheric Plasmas, ed. P.~{Heinzel},
  I.~{Dorotovi{\v c}}, \& R.~J. {Rutten}, 253

\bibitem[{{Pariat} {et~al.}(2012{\natexlab{b}}){Pariat}, {Schmieder}, {Masson},
  \& {Aulanier}}]{2012EAS....55..115P}
{Pariat}, E., {Schmieder}, B., {Masson}, S., \& {Aulanier}, G.
  2012{\natexlab{b}}, in EAS Publications Series, Vol.~55, EAS Publications
  Series, ed. M.~{Faurobert}, C.~{Fang}, \& T.~{Corbard}, 115--124

\bibitem[{{Qiu} {et~al.}(2000){Qiu}, {Ding}, {Wang}, {Denker}, \&
  {Goode}}]{2000ApJ...544L.157Q}
{Qiu}, J., {Ding}, M.~D., {Wang}, H., {Denker}, C., \& {Goode}, P.~R. 2000,
  \apjl, 544, L157

\bibitem[{{Roy}(1973)}]{1973SoPh...28...95R}
{Roy}, J.~R. 1973, \solphys, 28, 95

\bibitem[{{Roy} \& {Leparskas}(1973)}]{1973SoPh...30..449R}
{Roy}, J.-R., \& {Leparskas}, H. 1973, \solphys, 30, 449

\bibitem[{{Rutten} {et~al.}(2013){Rutten}, {Vissers}, {Rouppe van der Voort},
  {S{\"u}tterlin}, \& {Vitas}}]{2013JPhCS.440a2007R}
{Rutten}, R.~J., {Vissers}, G.~J.~M., {Rouppe van der Voort}, L.~H.~M.,
  {S{\"u}tterlin}, P., \& {Vitas}, N. 2013, Journal of Physics Conference
  Series, 440, 012007

\bibitem[{{Scharmer} {et~al.}(2003{\natexlab{a}}){Scharmer}, {Bjelksjo},
  {Korhonen}, {Lindberg}, \& {Petterson}}]{2003SPIE.4853..341S}
{Scharmer}, G.~B., {Bjelksjo}, K., {Korhonen}, T.~K., {Lindberg}, B., \&
  {Petterson}, B. 2003{\natexlab{a}}, in Society of Photo-Optical
  Instrumentation Engineers (SPIE) Conference Series, Vol. 4853, Society of
  Photo-Optical Instrumentation Engineers (SPIE) Conference Series, ed. S.~L.
  {Keil} \& S.~V. {Avakyan}, 341--350

\bibitem[{{Scharmer} {et~al.}(2003{\natexlab{b}}){Scharmer}, {Dettori},
  {L{\"o}fdahl}, \& {Shand}}]{2003SPIE.4853..370S}
{Scharmer}, G.~B., {Dettori}, P.~M., {L{\"o}fdahl}, M.~G., \& {Shand}, M.
  2003{\natexlab{b}}, in Society of Photo-Optical Instrumentation Engineers
  (SPIE) Conference Series, Vol. 4853, Society of Photo-Optical Instrumentation
  Engineers (SPIE) Conference Series, ed. S.~L. {Keil} \& S.~V. {Avakyan},
  370--380

\bibitem[{{Scharmer} {et~al.}(2008){Scharmer}, {Narayan}, {Hillberg}, {de la
  Cruz Rodr{\'{\i}}guez}, {L{\"o}fdahl}, {Kiselman}, {S{\"u}tterlin}, {van
  Noort}, \& {Lagg}}]{2008ApJ...689L..69S}
{Scharmer}, G.~B., {Narayan}, G., {Hillberg}, T., {et~al.} 2008, \apjl, 689,
  L69

\bibitem[{{Schmieder} {et~al.}(2004){Schmieder}, {Rust}, {Georgoulis},
  {D{\'e}moulin}, \& {Bernasconi}}]{2004ApJ...601..530S}
{Schmieder}, B., {Rust}, D.~M., {Georgoulis}, M.~K., {D{\'e}moulin}, P., \&
  {Bernasconi}, P.~N. 2004, \apj, 601, 530

\bibitem[{{Schnerr} {et~al.}(2011){Schnerr}, {de La Cruz Rodr{\'{\i}}guez}, \&
  {van Noort}}]{2011A&A...534A..45S}
{Schnerr}, R.~S., {de La Cruz Rodr{\'{\i}}guez}, J., \& {van Noort}, M. 2011,
  \aap, 534, A45

\bibitem[{{Severny}(1968)}]{1968mmsf.conf...71S}
{Severny}, A.~B. 1968, in Mass Motions in Solar Flares and Related Phenomena,
  ed. Y.~{Oehman}, 71

\bibitem[{{Shibata} {et~al.}(1982){Shibata}, {Nishikawa}, {Kitai}, \&
  {Suematsu}}]{1982SoPh...77..121S}
{Shibata}, K., {Nishikawa}, T., {Kitai}, R., \& {Suematsu}, Y. 1982, \solphys,
  77, 121

\bibitem[{{Shibata} {et~al.}(2007){Shibata}, {Nakamura}, {Matsumoto}, {Otsuji},
  {Okamoto}, {Nishizuka}, {Kawate}, {Watanabe}, {Nagata}, {UeNo}, {Kitai},
  {Nozawa}, {Tsuneta}, {Suematsu}, {Ichimoto}, {Shimizu}, {Katsukawa},
  {Tarbell}, {Berger}, {Lites}, {Shine}, \& {Title}}]{2007Sci...318.1591S}
{Shibata}, K., {Nakamura}, T., {Matsumoto}, T., {et~al.} 2007, Science, 318,
  1591

\bibitem[{{Shimizu} {et~al.}(2002){Shimizu}, {Shine}, {Title}, {Tarbell}, \&
  {Frank}}]{2002ApJ...574.1074S}
{Shimizu}, T., {Shine}, R.~A., {Title}, A.~M., {Tarbell}, T.~D., \& {Frank}, Z.
  2002, \apj, 574, 1074

\bibitem[{{Shine} {et~al.}(1994){Shine}, {Title}, {Tarbell}, {Smith}, {Frank},
  \& {Scharmer}}]{1994ApJ...430..413S}
{Shine}, R.~A., {Title}, A.~M., {Tarbell}, T.~D., {et~al.} 1994, \apj, 430, 413

\bibitem[{{Socas-Navarro} {et~al.}(2006){Socas-Navarro}, {Mart{\'{\i}}nez
  Pillet}, {Elmore}, {Pietarila}, {Lites}, \& {Manso
  Sainz}}]{2006SoPh..235...75S}
{Socas-Navarro}, H., {Mart{\'{\i}}nez Pillet}, V., {Elmore}, D., {et~al.} 2006,
  \solphys, 235, 75

\bibitem[{{van Noort} {et~al.}(2005){van Noort}, {Rouppe van der Voort}, \&
  {L{\"o}fdahl}}]{2005SoPh..228..191V}
{van Noort}, M., {Rouppe van der Voort}, L., \& {L{\"o}fdahl}, M.~G. 2005,
  \solphys, 228, 191

\bibitem[{{Vissers} \& {Rouppe van der Voort}(2012)}]{2012ApJ...750...22V}
{Vissers}, G., \& {Rouppe van der Voort}, L. 2012, \apj, 750, 22

\bibitem[{{Watanabe} {et~al.}(2012){Watanabe}, {Bellot Rubio}, {de la Cruz
  Rodr{\'{\i}}guez}, \& {Rouppe van der Voort}}]{2012ApJ...757...49W}
{Watanabe}, H., {Bellot Rubio}, L.~R., {de la Cruz Rodr{\'{\i}}guez}, J., \&
  {Rouppe van der Voort}, L. 2012, \apj, 757, 49

\bibitem[{{Watanabe} {et~al.}(2011){Watanabe}, {Vissers}, {Kitai}, {Rouppe van
  der Voort}, \& {Rutten}}]{2011ApJ...736...71W}
{Watanabe}, H., {Vissers}, G., {Kitai}, R., {Rouppe van der Voort}, L., \&
  {Rutten}, R.~J. 2011, \apj, 736, 71

\bibitem[{{Watanabe} {et~al.}(2008){Watanabe}, {Kitai}, {Okamoto}, {Nishida},
  {Kiyohara}, {Ueno}, {Hagino}, {Ishii}, \& {Shibata}}]{2008ApJ...684..736W}
{Watanabe}, H., {Kitai}, R., {Okamoto}, K., {et~al.} 2008, \apj, 684, 736

\bibitem[{{Yi} \& {Molowny-Horas}(1995)}]{1995A&A...295..199Y}
{Yi}, Z., \& {Molowny-Horas}, R. 1995, \aap, 295, 199

\bibitem[{{Yokoyama} \& {Shibata}(1995)}]{1995Natur.375...42Y}
{Yokoyama}, T., \& {Shibata}, K. 1995, \nat, 375, 42

\bibitem[{{Zachariadis} {et~al.}(1987){Zachariadis}, {Alissandrakis}, \&
  {Banos}}]{1987SoPh..108..227Z}
{Zachariadis}, T.~G., {Alissandrakis}, C.~E., \& {Banos}, G. 1987, \solphys,
  108, 227

\end{thebibliography}

\end{document}